\documentclass[article,superscriptaddress,aps,twocolumn,pra,longbibliography]{revtex4-1}
\usepackage{dcolumn}
\usepackage{bm}
\usepackage{epsfig}
\usepackage{graphicx}
\usepackage{amssymb,amsmath,amsbsy,amsgen,amsfonts,amsthm,mathtools}    
\usepackage{mathrsfs}
\usepackage{latexsym}
\usepackage{array}
\usepackage{color}
\usepackage{amstext}
\allowdisplaybreaks[1]
\usepackage{txfonts}
\usepackage{pifont}
\usepackage{verbatim}
\usepackage{hyperref}

\usepackage{epstopdf} 
\usepackage{todonotes} 

\newcommand{\abs}[1]{\left\vert #1\right\vert}

\newcommand{\ket}[1]{\left\vert{#1}\right\rangle}

\DeclareSymbolFont{symbols}{OMS}{cmsy}{m}{n}

\begin{document}
\title{
Using states with a large photon number variance to increase quantum Fisher information in single-mode phase estimation}

\author{Changhyoup Lee}
\email{changhyoup.lee@gmail.com}
\affiliation{Institute of Theoretical Solid State Physics, Karlsruhe Institute of Technology, 76131 Karlsruhe, Germany}

\author{Changhun Oh}
\affiliation{Center for Macroscopic Quantum Control, Department of Physics and Astronomy, Seoul National University, Seoul 08826, Korea}

\author{Hyunseok Jeong}
\affiliation{Center for Macroscopic Quantum Control, Department of Physics and Astronomy, Seoul National University, Seoul 08826, Korea}

\author{Carsten Rockstuhl}
\affiliation{Institute of Theoretical Solid State Physics, Karlsruhe Institute of Technology, 76131 Karlsruhe, Germany}
\affiliation{Institute of Nanotechnology, Karlsruhe Institute of Technology, 76021 Karlsruhe, Germany}

\author{Su-Yong~Lee}
\affiliation{Quantum Physics Technology Directorate, Agency for Defense Development, Daejeon, Korea}
\date{\today}

\begin{abstract}
When estimating the phase of a single mode, the quantum Fisher information for a pure probe state is proportional to the photon number variance of the probe state. In this work, we point out particular states that offer photon number distributions exhibiting a large variance, which would help to improve the local estimation precision. These theoretical examples are expected to stimulate the community to put more attention to those states that we found, and to work towards their experimental realization and usage in quantum metrology.
\end{abstract}



\maketitle

\section{Introduction}
Finding an optimal combination of an input state and a measurement setup is one of the key issues in quantum metrology, by which quantum enhancement can be maximized~\cite{Giovannetti2011}. On the one hand, the optimality of a measurement setting is assessed by comparing the Fisher information for a chosen setting with the quantum Fisher information (QFI) that would be obtained by an optimal setting, given parameter encoding and a probe state~\cite{Paris2009, Oh2019a}. The optimality of a probe state, on the other hand, can be addressed by maximizing the QFI given a parameter encoding~\cite{Dorner2009}. The aforementioned approaches apply to various parameter estimation problems.

Much attention has been paid on identifying optimal quantum states in a variety of quantum metrological applications. The attention has been triggered because the key mechanism leading to quantum enhancement can often be understood as the non-classicality of the probe state~\cite{Giovannetti2006,Giovannetti2011,Kwon2019}. For example, in single-mode loss parameter estimation, the photon number state having no uncertainty in the intensity is known to be the optimal state, providing the maximal quantum enhancement~\cite{Adesso2009,Nair2018}. In phase parameter estimation, it is known that the squeezed vacuum state reaches the QFI scaled with~$N^2$~\cite{Monras2006}, leading to a Heisenberg scaling of~$N^{-1}$ in precision, where $N$ is the average photon number of the probe state. However, the squeezed vacuum state is not the theoretical optimal state that maximizes the QFI in single-mode phase estimation as we will discuss through this work. 

Various fiducial photon number distributions have so far been considered as candidates to achieve quantum enhancement in single-mode phase estimation. Examples include the SSW state~\cite{Shapiro1989}, the SS state~\cite{Shapiro1991}, Dowling's model~\cite{Dowling1991}, the small peak model~\cite{Braunstein1994b, Rivas2012}. These states are respectively written in the photon number state basis $\{\ket{n}\}$ by
\begin{align}
\ket{\psi_\text{SSW}}&=\frac{1}{{\cal N}}\sum_{n=0}^{M}\frac{1}{n+1}\ket{n},\\
\ket{\psi_\text{SS}}&=\frac{1}{{\cal N}}\sum_{n=0}^{M}\frac{1}{n+z}\ket{n},\\
\ket{\psi_\text{Dowling}}&=\frac{1}{{\cal N}}\sum_{n=0}^{\infty}\frac{e^{-n/\eta}}{n+z}\ket{n},\\
\ket{\psi_\text{SMP}}&=\sqrt{1-a}\ket{0}+\sqrt{a}\ket{\pi},\label{SMP}
\end{align}
where the ${\cal N}$'s correspond to normalization factors, $z$ is a positive constant, $\eta$ is a smooth cutoff, $0\le a\le 1$, and $\ket{\pi}$ is orthogonal to the vacuum. Different approaches have been employed to show the advantages of such states in phase estimation. 

In this work, we begin with the appreciation that the QFI for the single-mode phase parameter estimation is proportional to the photon number variance of the probe state and sets the lower bound in the precision through the quantum Cram{\'e}r-Rao inequality~\cite{Braunstein1994a, Braunstein1996}. This implies that the probe state with the maximal photon number variance would possibly be the theoretical optimal state for single-mode phase estimation. Here, we aim to introduce, while leaving the proof of the achievability of the quantum Cram{\'e}r-Rao bound to future studies~\cite{Rubio2018, Rubio2019}, fiducial quantum states that have the maximum, or at least a larger photon number variance than that available with the squeezed vacuum state -- the paradigmatic state known to be useful for quantum phase estimation. We distinguish the scenarios when the photon number probability distribution is either bounded or unbounded, i.e., defined within a finite or an infinite domain~\cite{James2006}. When considering bounded distributions, we show that the theoretical optimal state with maximum photon number variance can indefinitely increase the QFI even for a fixed average photon number~$N$. When considering unbounded distributions, we show that one can achieve not only the Heisenberg scaling using other quantum states than the squeezed vacuum state, but also sub-Heisenberg scaling by a particular photon number statistics without relying on any nonlinear effects. Here, the sub-Heisenberg scaling manifests in terms of the average photon number $N$ and might mislead to conclude that it violates the fundamental Heisenberg limit. More details on that can be found in the relevant debates, which have been devoted over the last decade~\cite{Bollinger1996, Yurke1986, Sanders1995, Ou1996, Zwierz2010, Luis2013a, Luis2013b, Anisimov2010}, followed by the conclusive proofs~\cite{Zwierz2010, Tsang2012, Giovannetti2012a, Giovannetti2012b, Berry2012, Hall2012a,  Hall2012b, Hall2012c, Jarzyna2015}. The latter showed that the overall scaling, while including the amount of resources required for obtaining a priori probability distribution of the parameter and the number of measurements required to achieve the asymptotic bound, is still Heisenberg scaling-limited. Nevertheless, the fiducial photon number distributions we introduce here would be useful for an operating regime of a parameter that is locally calibrated in advance, so the identification of minute changes of the parameter is only of interest. That is, fortunately, often the case, e.g., for plasmonic sensors~\cite{Lee2017,Lee2018} or phase tracking~\cite{Yonezawa2012}. In such cases, the validity of the quantum Cram{\'e}r-Rao bound can be investigated in terms of the required minimum number of measurements and the minimum prior knowledge of the parameter~\cite{Rubio2018, Rubio2019}.

The theoretical states we discuss in this work have rarely been experimentally realized so far~\cite{McCormick2019}, but we expect more states will be implemented in the future. It would require the development of quantum technology geared towards engineering states with photon number statistics on demand. Recently, an arbitrary photon number statistics has been shown to be producible with current technology through quantum optical circuits being optimized for a target photon number statistics~\cite{DellAnno2006, Bimbard2010, Nichols2018, Arrazola2019}. Having the ability to prepare such quantum states unlocks their use for various purposes in quantum applications~\cite{ODriscoll2019}. Therefore, the purpose of this work is to lay out exotic photon number distributions in order to trigger experimental efforts along these lines.


\section{Phase estimation}
For a parameter-encoded pure state $\ket{\psi_\phi}=e^{i\phi\hat{G}}\ket{\psi_\text{in}}$, where $\hat{G}$ denotes a generator encoding a parameter $\phi$, the QFI can be calculated by $H= 4\langle (\Delta \hat{G})^{2}\rangle$~\cite{Braunstein1994a, Braunstein1996}, where $\langle (\Delta \hat{O})^2\rangle=\langle \hat{O}^2\rangle-\langle \hat{O}\rangle^2$ for an operator $\hat{O}$ and the expectation value is calculated for $\ket{\psi_\text{in}}$. The QFI sets the lower bound to the mean-squared-error of estimate when considering an unbiased estimator, given by the quantum Cram\'{e}r-Rao inequality written as
\begin{align}
\Delta\phi \ge\frac{1}{\sqrt{\nu H}},
\label{QCRI}
\end{align}
where $\Delta\phi$ is the root-mean-squared-error, interpreted as the estimation error or precision, and $\nu$ denotes the number of repetitions of measurement. This bound, called quantum Cram{\'e}r-Rao bound, is known to be achievable in the asymptotic limit $\nu\rightarrow\infty$.

For a single-mode phase parameter encoding, $\hat{G}=\hat{a}^{\dagger}\hat{a}$, so that the QFI is given by
\begin{align}
H=4\langle (\Delta \hat{n})^{2}\rangle 
\label{QFI}
\end{align}
where $\hat{n}=\hat{a}^{\dagger}\hat{a}$. This clearly indicates that a probe state $\ket{\psi_\text{in}}$ with a maximum photon number variance leads to the maximal QFI. The importance of the photon number fluctuation for phase estimation has been addressed~\cite{Hofmann2009, Hyllus2010}. In consequence, the maximum photon number variance leads to the greatest quantum enhancement over the standard quantum limit (SQL), i.e., $\Delta \phi$ scaled with $N^{-1/2}$~\cite{Leonhardt1995}. Such scaling is the optimal scaling that can be obtained when only classical resources are used~\cite{Pirandola2018}. Therefore, it is of utmost importance to identify quantum states with a maximum photon number variance.

To set the stage before looking for particular photon number distributions, let us consider a few of paradigmatic states that have often been considered for phase estimation. The first one is a coherent state $\ket{\alpha}$ of light, for which $H_\text{coh}=4N$, where the average photon number is $N=\abs{\alpha}^2$~\cite{Monras2006}. $H_\text{coh}$ is regarded as the classical benchmark in single-mode phase estimation, i.e., the SQL. Another example is a squeezed vacuum state written as $\ket{\xi}=\sum_n c_n \ket{2n}$ where $c_n=(-e^{i\theta_\text{s}}\tanh r)^{n}\sqrt{(2n)!}/\left(2^n n!\sqrt{\cosh r}\right)$ with the squeezing parameters $r$ and $\theta_\text{s}$. For the squeezed vacuum state, the QFI reads as~\cite{Monras2006}
\begin{align}
H_\text{sq}=8(N^2+N),
\label{QFI_sq}
\end{align}
where the average photon number is given as $N=\sinh^{2}r$. It is clear that $H_\text{sq}$ exhibits a Heisenberg scaling, which suggests that $\Delta \phi$ scales with $N^{-1}$ [see Eq.~\eqref{QCRI}]. In particular, one can see that $H_\text{sq}\approx 504.89$ for state-of-the-art squeezed state of 15 dB-squeezing as recently reported~\cite{Vahlbruch2016}, approximately corresponding to $r\approx1.73$ (i.e., $N\approx 7.46$) while ignoring the thermal photon contribution for simplicity despite its practical significance studied in Refs.~\cite{Aspachs2009, Safranek2016, Oh2019a, Oh2019b}.

In the next sections, we look for quantum states with maximum photon number variance, or at least larger than that of the squeezed vacuum state, which consequently further increases the QFI in Eq.~\eqref{QFI} as compared to $H_\text{sq}$ of Eq.~\eqref{QFI_sq}. To this end, we distinguish two types of discrete probability distributions $p(n)$ for photon number statistics of a single-mode probe state: a bounded photon number distribution that is defined within a finite domain $n\in[m,M]$ with integers $m<M$ and an unbounded photon number distribution that is defined in an infinite domain $n\in[0,\infty)$.

\section{Bounded photon number distributions}
For the sake of generality, let us consider an arbitrary superposition of photon number states in a range from $m$ to $M$ photons, written as
\begin{align}
\ket{\psi_\text{b}}=\sum_{n=m}^{M}\sqrt{p(n)}e^{i\theta(n)} \ket{n},
\end{align}
where the photon number distribution $p(n)$ is bounded by the minimum $m$ and the maximum $M$, i.e., $p(n)=0$ for $n< m$ and $n> M$. The phase distribution $\theta(n)$ plays an important role in preparing an optimal measurement setting in practice, which depends on both $\theta(n)$ and $\phi$ being estimated. The phases, however, can be dismissed in this work since we focus on the error bound given by the QFI. This means that the optimal measurement setting assumed to be chosen accommodates the phases, leaving only the dependence of $p(n)$ in Eq.~\eqref{QFI}.
One can find that the variance of such bounded probability distribution $p(n)$ is upper bounded by Popoviciu's inequality~\cite{Popoviciu1935}, given as
\begin{align}
\langle (\Delta \hat{n})^2 \rangle \le \frac{1}{4}(M-m)^2,
\label{Popoviciu_Inequality}
\end{align}
where the equality holds when $p(m)=p(M)=1/2$. This implies that for the given minimum $m$ and maximum $M$, a balanced superposition of $m$ and $M$ photons provides the maximal QFI according to Eq.~\eqref{QFI}. The QFI is thus written as $H=4(M-N)^2$ with $N=(m+M)/2$ being the average photon number. For a fixed $N$, the maximal QFI is obtained when $m=0$, which is obvious, for which $H=4N^2$, clearly showing the Heisenberg scaling, but still smaller than $H_\text{sq}$ in Eq.~\eqref{QFI_sq}. The bound on $\Delta\phi$ associated Popoviciu's inequality indicates that the Heisenberg scaling is the maximal scaling when the photon number distribution is bounded. 

A stronger inequality than Eq.~\eqref{Popoviciu_Inequality} exists, called the Bhatia–Davis inequality~\cite{Bhatia-Davis2000}, which is written as
\begin{align}
\langle (\Delta \hat{n})^2 \rangle \le (M-N)(N-m),
\label{Bhatia_Davis_Inequality}
\end{align}
where the equality holds when $p(n)=1-a$ and $p(M)=a$ for an arbitrary weight factor of $a$ that determines the average photon number $N=(1-a)m+a M$. When $a=1/2$, the Bhatia-Davis inequality of Eq.~\eqref{Bhatia_Davis_Inequality} becomes the Popoviciu inequality of Eq.~\eqref{Popoviciu_Inequality}. The Bhatia-Davis inequality suggests to consider an arbitrary superposition state of $m$ and $M$ photons, which we call the m\&M state throughout this work. The m\&M state can be written as
\begin{align}
\ket{\psi_{\text{m\&M}}}=\sqrt{1-a}\ket{m}+\sqrt{a}\ket{M}.
\label{m&Mstate}
\end{align}
This leads to the QFI of the form
\begin{align}
H_{\text{m\&M}}
=4a(1-a) (M-m)^2.
\label{QFI_m&M}
\end{align}
It is clear that $H_{\text{m\&M}}$ depends on the difference $(M-m)$ and takes on the maximum when $a=1/2$ for given $m$ and $M$, the case satisfying the equality of Popoviciu's inequality. To compare the QFIs for a fixed $N$, let us set $a=(N-m)/(M-m)$ which keeps $N$ unchanged for any $m$ and $M$, so that Eq.~\eqref{m&Mstate} is rewritten by
\begin{align}
\ket{\psi_{\text{m\&M}}}=\sqrt{\frac{M-N}{M-m}}\ket{m}+\sqrt{\frac{N-m}{M-m}}\ket{M},
\label{m&Mstate2}
\end{align}
and Eq.~\eqref{QFI_m&M} becomes
\begin{align}
H_{\text{m\&M}}
=4(M-N)(N-m).
\label{QFI_m&M2}
\end{align}
Note that $N$ is fixed in Eq.~\eqref{QFI_m&M2} regardless of the values of $m$ and $M$ although Eq.~\eqref{QFI_m&M2} seems directly obtainable from Eq.~\eqref{Bhatia_Davis_Inequality} where $N$ definitely depends on $m$ and $M$. It is interesting to see that in the limit $M\gg N$, one obtains $H_{\text{m\&M}}\approx 4 M(N-m)$, which can be arbitrarily increased by increasing $M$ while keeping $N$ fixed. 

Equation~\eqref{QFI_m&M2} indicates that the QFI increases with increasing the maximum $M$ and decreasing the minimum $m$ for a fixed $N$. So let us set $m=0$, for which the m\&M state of Eq.~\eqref{m&Mstate2} becomes the 0\&M state, i.e., $\ket{\psi_{\text{0\&M}}}=\sqrt{(M-N)/M}\ket{0}+\sqrt{N/M}\ket{M}$, for which $H_{\text{0\&M}}=4N(M-N)$. Therefore, the 0\&M state is the optimal state and $H_{\text{0\&M}}$ is the upper bound for the QFI within the class of the states having a bounded photon number distribution. The 0\&M state has been considered as the so-called ON states in the context of quantum computation~\cite{Sabapathy2018} and a few schemes for its experimental generation have been proposed~\cite{Yukawa2013,Arrazola2019}. The 0\&M state has already been discussed as the state showing an arbitrarily large QFI in single-mode phase estimation~\cite{Berry2012,Luis2017}, but here we prove, by using the Bhatia-Davis inequality of Eq.~\eqref{Bhatia_Davis_Inequality}, that the 0\&M state is the theoretical optimal state exhibiting the maximum photon number variance among the states with bounded photon number distributions. 

The 0\&M state can be categorized as the small peak model of Eq.~\eqref{SMP}. In general, the QFI for the small peak model is given as $H_\text{SMP}=4 N(N_\pi-N)+4(\Delta n_\pi)^2 N/N_\pi$, where $N_\pi$ is the average photon number of the state $\ket{\pi}$ and $(\Delta n_\pi)^2$ denotes its variance. The small peak model is able to attain an arbitrarily large QFI by increasing either $N_\pi$ or $(\Delta n_\pi)^2$, while keeping $N$ fixed. The particular case $\ket{\pi}=\ket{\xi}$ has been discussed in Ref.~\cite{Rivas2012}, followed by the review in Ref.~\cite{Berry2012}.

\begin{figure}[t]
\centering
\includegraphics[width=0.4\textwidth]{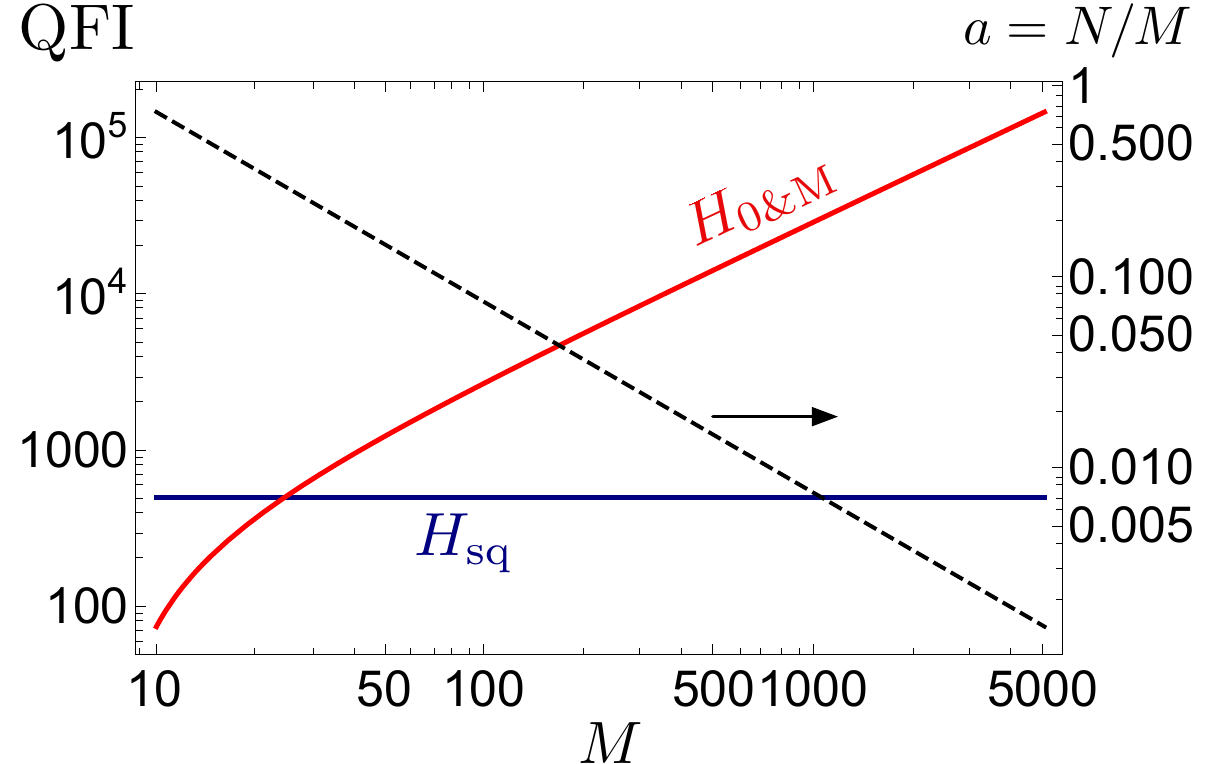}
\caption{
The QFI for the 0\&M state, $H_\text{0\&M}$ (red curve), can be arbitrarily increased with $M$, in comparison with $H_\text{sq}$ (blue curve) for $N=7.46$ as an example. The weight factor $a$ (dashed curve) is set to keep the average photon number $N$ unchanged while varying $M$.}
\label{Fig:m&Mstate}
\end{figure}

In comparison with Eq.~\eqref{QFI_sq}, for $N=7.46$ considered in state-of-the-art squeezed vacuum state, one can achieve higher QFI than $H_\text{sq}$ with the 0\&M state when $M\ge25$ (corresponding to $a\lesssim0.3$), resulting in $H_\text{0\&M}\gtrsim 523.39$. 
Figure~\ref{Fig:m&Mstate} shows the behaviors of $H_\text{0\&M}$ (see red curve) and $a=N/M$ (see dashed curve) with varying $M$ for $N=7.46$. Note that $H_\text{0\&M}$ in the order of $10^5$ can be theoretically attained by increasing $M$ even when $N$ is fixed. 
The 0\&M state has been realized up to $M=18$ in the harmonic motion of a single trapped ion~\cite{McCormick2019}, and the states with higher $M$ can also be realized in quantum optical circuits with current technology~\cite{Nichols2018, Arrazola2019, ODriscoll2019}.

\section{Unbounded photon number distributions}
When a probability distribution is defined in an infinite domain, i.e., unbounded, there exits an infinite number of degrees of freedom to characterize types of unbounded probability distribution. Therefore, the analysis for unbounded photon number distributions would not be as simple as the bounded case. Instead, we investigate here a few special probability distributions, which lead to intriguing behaviors in single-mode phase estimation.

\subsection{Heisenberg scaling in the local precision}
As mentioned above, the squeezed vacuum state enables the Heisenberg scaling of $N^{-1}$ in $\Delta\phi$. It is interesting to see that there exist other types of photon number statistics, leading to the Heisenberg scaling in phase estimation. Below, let us look at some of them as examples. 

Consider the probe state with the photon number distribution given as 
\begin{align}
p_\text{G}(n)=\mu(1-\mu)^{n},
\end{align}
for $\mu\in(0,1)$. This is called the geometric distribution and $p_\text{G}(n)$ is the probability of $n+1$ Bernoulli trials required to get the first success with success probability $\mu$. It possesses the average photon number of $N=(1-\mu)/\mu$ and the variance of
\begin{align}
\langle (\Delta \hat{n})^2\rangle_\text{G} =N^2+N.
\end{align}
This clearly exhibits the Heisenberg scaling through Eq.~\eqref{QFI}, i.e., scaling of $N^{-1}$ in $\Delta \phi$, although a little worse than the case using a squeezed vacuum state due to the absence of the factor of 2. 

A generalization of the geometric distribution, called the negative binomial distribution, can also be considered, written by
\begin{align}
p_\text{NB}(n)=\binom{n+\eta-1}{n}\mu^n(1-\mu)^\eta
\end{align}
for $\mu\in(0,1)$ and $\eta>0$. In this case, the average photon number is given by $N=\mu\eta/(1-\mu)$ while the variance takes the form of
\begin{align}
\langle (\Delta \hat{n})^2\rangle_\text{NB} =N^2+\frac{\mu \eta (1-\mu\eta)}{(1-\mu)^2}.
\end{align}
Note that the second term is positive only when $\mu\eta<1$, for which the Heisenberg scaling is achieved. When $\mu\eta>1$, on the other hand, a worse scaling than the Heisenberg scaling is obtained. It can be shown that the states with $p_\text{NB}(n)$ significantly outperforms the case using a squeezed vacuum state when $\mu\eta<1$ and $\mu\approx1$.

As another example, consider the probe state with the photon number distribution given as
\begin{align}
p_\text{L}(n)=
\begin{cases}
0      & \text{for~} n=0, \\
\frac{-1}{\ln(1-\mu)}\frac{\mu^n}{n}  & \text{for~} n\ge1,
\end{cases}
\end{align}
for $\mu\in(0,1)$. This is called the logarithmic distribution and has been used to model relative species abundance~\cite{Johnson2005}. It exhibits the average photon number of ${N=-\mu/(1-\mu)\ln(1-\mu)}$ and the variance of 
\begin{align}
\langle (\Delta \hat{n})^2\rangle_\text{L} 
=N^2+\frac{-\mu[2\mu +\ln(1-\mu)]}{(1-\mu)^2 [ \ln(1-\mu) ]^2}.
\end{align}
Here, the second term plays an important role in determining a further improvement when compared to the case using a squeezed vacuum state. The second term is negative when $\mu<\mu_\text{c}$. It crosses zero to be positive at $\mu=\mu_\text{c}$, and increases to diverge when increasing $\mu$ further, where $\mu_c\approx0.7968$ is the solution of $2\mu+\ln[1-\mu]=0$.
One can see that the corresponding QFI is less than $H_\text{sq}$ for $\mu\ll1$ (i.e., $N\ll1$), but outperforms $H_\text{sq}$ when $\mu\approx 1$ (i.e., $N\gg1$). 

\subsection{Sub-Heisenberg scaling in the local precision}
The Heisenberg scaling of $N^{-1}$ in $\Delta \phi$ is considered as the ultimate scaling in quantum parameter estimation, often called the Heisenberg limit.  It has been shown that a sub-Heisenberg scaling~\footnote{In the literature, the terms ``super-Heisenberg scaling'' and ``sub-Heisenberg scaling'' have interchangeably used to denote the same limit~\cite{Boixo2008, Woolley2008, Rams2018, Roy2019}.} of $N^{-s}$ with $s>1$ is achievable through nonlinear effects arising in many-body systems~\cite{Boixo2007,Boixo2008,Choi2008,Roy2008,Woolley2008,Napolitano2010,Rams2018}. The latter has been demonstrated with a nonlinear atomic ensemble~\cite{Napolitano2011}. Here we show that a similar sub-Heisenberg scaling can also be achieved by particular photon number statistics of a single-mode state of light, but requiring neither nonlinearity nor many-body systems. Note that such alluring results do not indicate that the Heisenberg limit can be beaten, but have been proved to be still limited by the Heisenberg scaling when appropriately accounting of all the resources needed to reach the error bound~\cite{Zwierz2010, Tsang2012, Giovannetti2012a, Giovannetti2012b, Berry2012, Hall2012a,  Hall2012b, Hall2012c, Jarzyna2015}. 

Consider the state with the photon number distribution given by
\begin{align}
p_\text{B}(n)=
\begin{cases}
0      & \text{for~} n=0, \\
\frac{e^{-\mu n}(\mu n)^{n-1}}{n!}  & \text{for~} n\ge1,
\end{cases}
\label{Borel_dist}
\end{align}
for $\mu\in[0,1]$. The distribution $p_\text{B}(n)$ is called Borel distribution~\cite{Borel1942, Tanner1961}, being observed in branching process and queueing theory~\cite{Otter1949,Haight1960}.
The distribution of Eq.~\eqref{Borel_dist} exhibits the average photon number of $N=1/(1-\mu)$ and the variance of 
\begin{align}
\langle (\Delta \hat{n})^2\rangle_\text{B} = \mu/(1-\mu)^3=N^2(N-1),
\end{align}
obviously leading to the QFI of $H_\text{B}=4N^2(N-1)$. Therefore, the probe state engineered with photon number distribution of $p_\text{B}(n)$ promises sub-Heisenberg scaling of $N^{-3/2}$ in $\Delta \phi$, being dominant in the limit $N\gg1$, i.e., when $\mu\approx1$. Again, note that it has been proven that sub-Heisenberg strategies are not so effective~\cite{Giovannetti2012a}, but would provide rather insignificant improvement when taking into account a priori knowledge about the parameter $\phi$.


\subsection{Indefinite scaling in the local precision}
Unlike the bounded probability distribution, there is no upper bound to the variance of the unbounded probability distribution. In other words, some probability distribution may have a diverging or even an infinite variance, arising from the feature of heavier tails than the exponential distribution~\cite{Foss2013}. One can consider distributions such as the Riemann-Zeta distribution, the Beta negative binomial, or the Yule-Simon distribution, all of which exhibit a diverging or an infinite variance of the photon number. Particularly, the Riemann Zeta distribution has already been considered as an interesting example showing an infinite QFI in two-mode schemes~\cite{Zhang2013}. These examples seem to provide the completely precise estimation, but it turned out that it is not the case (see more detailed discussion in Ref.~\cite{Berry2012}). 

\section{Conclusion}
We have identified particular fiducial photon number distributions of a single-mode probe state, which maximize the QFI and would possibly be useful for the local phase estimation. Considering the case that the photon number distribution is bounded, we have provided the proof that the theoretical optimal state is the 0\&M state, indefinitely increasing the QFI and consequently reducing the local estimation error of $\Delta \phi$ in the asymptotic limit of the number of measurements $\nu\rightarrow\infty$. For the case that the photon number distribution is unbounded, on the other hand, we have discussed several particular photon number statistics which show Heisenberg scaling and sub-Heisenberg scaling without requiring nonlinear effects. The states discussed in this work have rarely been experimentally realized~\cite{McCormick2019}, but state-of-the-art quantum state engineering technique would enable the generation of an arbitrary photon number superposition via quantum circuit optimization~\cite{Nichols2018, Arrazola2019, ODriscoll2019}.
In the scenario when a priori probability distribution of the parameter is unknown and the number of measurements is limited, those states may not be useful since they are still Heisenberg-scaling limited with $N_\text{tot}=N\nu$, the total average number of photons being used. It has been shown that the strong Heisenberg limit written as $\Delta\phi_\text{s}\propto1/N_\text{tot}$~\cite{Pezze2008, Giovannetti2004, Giovannetti2011, Tsang2012, Giovannetti2012b, Berry2012, Hall2012a, Giovannetti2012a, Hall2012c, Hall2012b} can never be beaten~\cite{Shapiro1989,Shapiro1991,Hradil1992a,Hradil1992b,Braunstein1992a, Braunstein1992b,Lane1993, Braunstein1994b, Anisimov2010, Rivas2012,Luis2013a,Luis2013b, Zhang2013, Demkowicz-Dobrzanski2012,Pezze2013}. However, when estimating the parameter in a local regime, the states we discussed would be able to provide the sub-Heisenberg scaling in principle. Furthermore, Luis recently showed through analytical and numerical examination that the weak Heisenberg limit~\cite{Pezze2015}, written as $\Delta \phi_\text{w}\propto1/\sqrt{\nu}N$, can be beaten by the 0\&M state with the prior information being updated without bias~\cite{Luis2017}. 

More rigorous analysis beyond the framework of the quantum Cram{\'e}r-Rao bound is necessary to see whether or not the states discussed in this work can beat at least the weak Heisenberg limit for practical purposes~\cite{Rubio2018, Rubio2019}. We leave similar investigation for unbounded photon number distributions as a future study. From a more fundamental perspective, the relation between the QFI and quantum coherence can be investigated for the states discussed in this work~\cite{Streltsov2017, Giorda2018, Tan2018, Kwon2018}.
From a practical perspective, on the other hand, the effect of loss or decoherence needs to be taken into account when the local precision is more rigorously examined. These subtle analyses are beyond the scope of this work, and so we leave them for future work. 
It would also be interesting to investigate other kinds of single-mode parameter estimation or multi-mode schemes. Particularly, in the Mach-Zehnder interferometer, useful states within the class of path-symmetric states have been discussed in terms of the QFI in Ref.~\cite{Lee2016}. One can generalize it to an arbitrary two-mode setting for full generality.

\section*{acknowledgments}
We acknowledge support by the KIT-Publication Fund of the Karlsruhe Institute of Technology.
S.-Y.~Lee was supported by a grant to Quantum Frequency Conversion Project funded by Defense Acquisition Program Administration and Agency for Defense Development. 
C.~Oh and H.~Jeong were supported by the National Research Foundation of Korea (NRF) through grants funded by the Ministry of Science and ICT (Grant Nos. 2019R1H1A3079890 and NRF-2019M3E4A1080074 ).

\bibliography{reference.bib}

\begin{thebibliography}{89}%
\makeatletter
\providecommand \@ifxundefined [1]{%
 \@ifx{#1\undefined}
}%
\providecommand \@ifnum [1]{%
 \ifnum #1\expandafter \@firstoftwo
 \else \expandafter \@secondoftwo
 \fi
}%
\providecommand \@ifx [1]{%
 \ifx #1\expandafter \@firstoftwo
 \else \expandafter \@secondoftwo
 \fi
}%
\providecommand \natexlab [1]{#1}%
\providecommand \enquote  [1]{``#1''}%
\providecommand \bibnamefont  [1]{#1}%
\providecommand \bibfnamefont [1]{#1}%
\providecommand \citenamefont [1]{#1}%
\providecommand \href@noop [0]{\@secondoftwo}%
\providecommand \href [0]{\begingroup \@sanitize@url \@href}%
\providecommand \@href[1]{\@@startlink{#1}\@@href}%
\providecommand \@@href[1]{\endgroup#1\@@endlink}%
\providecommand \@sanitize@url [0]{\catcode `\\12\catcode `\$12\catcode
  `\&12\catcode `\#12\catcode `\^12\catcode `\_12\catcode `\%12\relax}%
\providecommand \@@startlink[1]{}%
\providecommand \@@endlink[0]{}%
\providecommand \url  [0]{\begingroup\@sanitize@url \@url }%
\providecommand \@url [1]{\endgroup\@href {#1}{\urlprefix }}%
\providecommand \urlprefix  [0]{URL }%
\providecommand \Eprint [0]{\href }%
\providecommand \doibase [0]{http://dx.doi.org/}%
\providecommand \selectlanguage [0]{\@gobble}%
\providecommand \bibinfo  [0]{\@secondoftwo}%
\providecommand \bibfield  [0]{\@secondoftwo}%
\providecommand \translation [1]{[#1]}%
\providecommand \BibitemOpen [0]{}%
\providecommand \bibitemStop [0]{}%
\providecommand \bibitemNoStop [0]{.\EOS\space}%
\providecommand \EOS [0]{\spacefactor3000\relax}%
\providecommand \BibitemShut  [1]{\csname bibitem#1\endcsname}%
\let\auto@bib@innerbib\@empty
\bibitem [{\citenamefont {Giovannetti}\ \emph {et~al.}(2011)\citenamefont
  {Giovannetti}, \citenamefont {Lloyd},\ and\ \citenamefont
  {Maccone}}]{Giovannetti2011}%
  \BibitemOpen
  \bibfield  {author} {\bibinfo {author} {\bibfnamefont {V.}~\bibnamefont
  {Giovannetti}}, \bibinfo {author} {\bibfnamefont {S.}~\bibnamefont {Lloyd}},
  \ and\ \bibinfo {author} {\bibfnamefont {L.}~\bibnamefont {Maccone}},\
  }\bibfield  {title} {\enquote {\bibinfo {title} {Advances in quantum
  metrology},}\ }\href@noop {} {\bibfield  {journal} {\bibinfo  {journal} {Nat.
  Photon.}\ }\textbf {\bibinfo {volume} {5}},\ \bibinfo {pages} {222} (\bibinfo
  {year} {2011})}\BibitemShut {NoStop}%
\bibitem [{\citenamefont {Paris}(2009)}]{Paris2009}%
  \BibitemOpen
  \bibfield  {author} {\bibinfo {author} {\bibfnamefont {M.~G.~A.}\
  \bibnamefont {Paris}},\ }\bibfield  {title} {\enquote {\bibinfo {title}
  {Quantum estimation for quantum technology},}\ }\href@noop {} {\bibfield
  {journal} {\bibinfo  {journal} {Int. J. Quantum Inf.}\ }\textbf {\bibinfo
  {volume} {7}},\ \bibinfo {pages} {125} (\bibinfo {year} {2009})}\BibitemShut
  {NoStop}%
\bibitem [{\citenamefont {Oh}\ \emph {et~al.}(2019)\citenamefont {Oh},
  \citenamefont {Lee}, \citenamefont {Rockstuhl}, \citenamefont {Jeong},
  \citenamefont {Kim}, \citenamefont {Nha},\ and\ \citenamefont
  {Lee}}]{Oh2019a}%
  \BibitemOpen
  \bibfield  {author} {\bibinfo {author} {\bibfnamefont {C.}~\bibnamefont
  {Oh}}, \bibinfo {author} {\bibfnamefont {C.}~\bibnamefont {Lee}}, \bibinfo
  {author} {\bibfnamefont {C.}~\bibnamefont {Rockstuhl}}, \bibinfo {author}
  {\bibfnamefont {H.}~\bibnamefont {Jeong}}, \bibinfo {author} {\bibfnamefont
  {J.}~\bibnamefont {Kim}}, \bibinfo {author} {\bibfnamefont {H.}~\bibnamefont
  {Nha}}, \ and\ \bibinfo {author} {\bibfnamefont {S.-Y.}\ \bibnamefont
  {Lee}},\ }\bibfield  {title} {\enquote {\bibinfo {title} {Optimal gaussian
  measurements for phase estimation in single-mode gaussian metrology},}\
  }\href@noop {} {\bibfield  {journal} {\bibinfo  {journal} {npj Quantum Inf.}\
  }\textbf {\bibinfo {volume} {5}},\ \bibinfo {pages} {10} (\bibinfo {year}
  {2019})}\BibitemShut {NoStop}%
\bibitem [{\citenamefont {Dorner}\ \emph {et~al.}(2009)\citenamefont {Dorner},
  \citenamefont {Demkowicz-Dobrzanski}, \citenamefont {Smith}, \citenamefont
  {Lundeen}, \citenamefont {Wasilewski}, \citenamefont {Banaszek},\ and\
  \citenamefont {Walmsley}}]{Dorner2009}%
  \BibitemOpen
  \bibfield  {author} {\bibinfo {author} {\bibfnamefont {U.}~\bibnamefont
  {Dorner}}, \bibinfo {author} {\bibfnamefont {R.}~\bibnamefont
  {Demkowicz-Dobrzanski}}, \bibinfo {author} {\bibfnamefont {B.~J.}\
  \bibnamefont {Smith}}, \bibinfo {author} {\bibfnamefont {J.~S.}\ \bibnamefont
  {Lundeen}}, \bibinfo {author} {\bibfnamefont {W.}~\bibnamefont {Wasilewski}},
  \bibinfo {author} {\bibfnamefont {K.}~\bibnamefont {Banaszek}}, \ and\
  \bibinfo {author} {\bibfnamefont {I.~A.}\ \bibnamefont {Walmsley}},\
  }\bibfield  {title} {\enquote {\bibinfo {title} {Optimal quantum phase
  estimation},}\ }\href@noop {} {\bibfield  {journal} {\bibinfo  {journal}
  {Phys. Rev. Lett.}\ }\textbf {\bibinfo {volume} {102}},\ \bibinfo {pages}
  {040403} (\bibinfo {year} {2009})}\BibitemShut {NoStop}%
\bibitem [{\citenamefont {Giovannetti}\ \emph {et~al.}(2006)\citenamefont
  {Giovannetti}, \citenamefont {Lloyd},\ and\ \citenamefont
  {Maccone}}]{Giovannetti2006}%
  \BibitemOpen
  \bibfield  {author} {\bibinfo {author} {\bibfnamefont {V.}~\bibnamefont
  {Giovannetti}}, \bibinfo {author} {\bibfnamefont {S.}~\bibnamefont {Lloyd}},
  \ and\ \bibinfo {author} {\bibfnamefont {L.}~\bibnamefont {Maccone}},\
  }\bibfield  {title} {\enquote {\bibinfo {title} {Quantum metrology},}\
  }\href@noop {} {\bibfield  {journal} {\bibinfo  {journal} {Phys. Rev. Lett.}\
  }\textbf {\bibinfo {volume} {96}},\ \bibinfo {pages} {010401} (\bibinfo
  {year} {2006})}\BibitemShut {NoStop}%
\bibitem [{\citenamefont {Kwon}\ \emph {et~al.}(2019)\citenamefont {Kwon},
  \citenamefont {Tan}, \citenamefont {Volkoff},\ and\ \citenamefont
  {Jeong}}]{Kwon2019}%
  \BibitemOpen
  \bibfield  {author} {\bibinfo {author} {\bibfnamefont {H.}~\bibnamefont
  {Kwon}}, \bibinfo {author} {\bibfnamefont {K.~C.}\ \bibnamefont {Tan}},
  \bibinfo {author} {\bibfnamefont {T.}~\bibnamefont {Volkoff}}, \ and\
  \bibinfo {author} {\bibfnamefont {H.}~\bibnamefont {Jeong}},\ }\bibfield
  {title} {\enquote {\bibinfo {title} {Nonclassicality as a quantifiable
  resource for quantum metrology},}\ }\href@noop {} {\bibfield  {journal}
  {\bibinfo  {journal} {Phys. Rev. Lett.}\ }\textbf {\bibinfo {volume} {122}},\
  \bibinfo {pages} {040503} (\bibinfo {year} {2019})}\BibitemShut {NoStop}%
\bibitem [{\citenamefont {Adesso}\ \emph {et~al.}(2009)\citenamefont {Adesso},
  \citenamefont {Dell’Anno}, \citenamefont {Siena}, \citenamefont
  {Illuminati},\ and\ \citenamefont {Souza}}]{Adesso2009}%
  \BibitemOpen
  \bibfield  {author} {\bibinfo {author} {\bibfnamefont {G.}~\bibnamefont
  {Adesso}}, \bibinfo {author} {\bibfnamefont {F.}~\bibnamefont {Dell’Anno}},
  \bibinfo {author} {\bibfnamefont {S.~De}\ \bibnamefont {Siena}}, \bibinfo
  {author} {\bibfnamefont {F.}~\bibnamefont {Illuminati}}, \ and\ \bibinfo
  {author} {\bibfnamefont {L.~A.~M.}\ \bibnamefont {Souza}},\ }\bibfield
  {title} {\enquote {\bibinfo {title} {Optimal estimation of losses at the
  ultimate quantum limit with non-gaussian states},}\ }\href@noop {} {\bibfield
   {journal} {\bibinfo  {journal} {Phys. Rev. A}\ }\textbf {\bibinfo {volume}
  {79}},\ \bibinfo {pages} {040305(R)} (\bibinfo {year} {2009})}\BibitemShut
  {NoStop}%
\bibitem [{\citenamefont {Nair}(2018)}]{Nair2018}%
  \BibitemOpen
  \bibfield  {author} {\bibinfo {author} {\bibfnamefont {R.}~\bibnamefont
  {Nair}},\ }\bibfield  {title} {\enquote {\bibinfo {title} {Quantum-limited
  loss sensing: Multiparameter estimation and bures distance between loss
  channels},}\ }\href@noop {} {\bibfield  {journal} {\bibinfo  {journal} {Phys.
  Rev. Lett.}\ }\textbf {\bibinfo {volume} {121}},\ \bibinfo {pages} {230801}
  (\bibinfo {year} {2018})}\BibitemShut {NoStop}%
\bibitem [{\citenamefont {Monras}(2006)}]{Monras2006}%
  \BibitemOpen
  \bibfield  {author} {\bibinfo {author} {\bibfnamefont {A.}~\bibnamefont
  {Monras}},\ }\bibfield  {title} {\enquote {\bibinfo {title} {Optimal phase
  measurements with pure gaussian states},}\ }\href@noop {} {\bibfield
  {journal} {\bibinfo  {journal} {Phys. Rev. A}\ }\textbf {\bibinfo {volume}
  {73}},\ \bibinfo {pages} {033821} (\bibinfo {year} {2006})}\BibitemShut
  {NoStop}%
\bibitem [{\citenamefont {Shapiro}\ \emph {et~al.}(1989)\citenamefont
  {Shapiro}, \citenamefont {Shepard},\ and\ \citenamefont
  {Wong}}]{Shapiro1989}%
  \BibitemOpen
  \bibfield  {author} {\bibinfo {author} {\bibfnamefont {J.~H.}\ \bibnamefont
  {Shapiro}}, \bibinfo {author} {\bibfnamefont {S.~R.}\ \bibnamefont
  {Shepard}}, \ and\ \bibinfo {author} {\bibfnamefont {N.~W.}\ \bibnamefont
  {Wong}},\ }\bibfield  {title} {\enquote {\bibinfo {title} {Ultimate quantum
  limits on phase measurement},}\ }\href@noop {} {\bibfield  {journal}
  {\bibinfo  {journal} {Phys. Rev. Lett}\ }\textbf {\bibinfo {volume} {62}},\
  \bibinfo {pages} {2377} (\bibinfo {year} {1989})}\BibitemShut {NoStop}%
\bibitem [{\citenamefont {Shapiro}\ and\ \citenamefont
  {Shepard}(1991)}]{Shapiro1991}%
  \BibitemOpen
  \bibfield  {author} {\bibinfo {author} {\bibfnamefont {J.~H.}\ \bibnamefont
  {Shapiro}}\ and\ \bibinfo {author} {\bibfnamefont {S.~R.}\ \bibnamefont
  {Shepard}},\ }\bibfield  {title} {\enquote {\bibinfo {title} {Quantum phase
  measurement: A system-theory perspective},}\ }\href@noop {} {\bibfield
  {journal} {\bibinfo  {journal} {Phys. Rev. A}\ }\textbf {\bibinfo {volume}
  {43}},\ \bibinfo {pages} {3795} (\bibinfo {year} {1991})}\BibitemShut
  {NoStop}%
\bibitem [{\citenamefont {Dowling}(1991)}]{Dowling1991}%
  \BibitemOpen
  \bibfield  {author} {\bibinfo {author} {\bibfnamefont {J.~P.}\ \bibnamefont
  {Dowling}},\ }\bibfield  {title} {\enquote {\bibinfo {title} {A quantum state
  of ultra-low phase noise},}\ }\href@noop {} {\bibfield  {journal} {\bibinfo
  {journal} {Opt. Commun.}\ }\textbf {\bibinfo {volume} {86}},\ \bibinfo
  {pages} {119} (\bibinfo {year} {1991})}\BibitemShut {NoStop}%
\bibitem [{\citenamefont {Braunstein}(1994)}]{Braunstein1994b}%
  \BibitemOpen
  \bibfield  {author} {\bibinfo {author} {\bibfnamefont {S.~L.}\ \bibnamefont
  {Braunstein}},\ }\bibfield  {title} {\enquote {\bibinfo {title} {Some limits
  to precision phase measurement},}\ }\href@noop {} {\bibfield  {journal}
  {\bibinfo  {journal} {Phys. Rev. A}\ }\textbf {\bibinfo {volume} {49}},\
  \bibinfo {pages} {69} (\bibinfo {year} {1994})}\BibitemShut {NoStop}%
\bibitem [{\citenamefont {Rivas}\ and\ \citenamefont {Luis}(2012)}]{Rivas2012}%
  \BibitemOpen
  \bibfield  {author} {\bibinfo {author} {\bibfnamefont {{\' A}.}~\bibnamefont
  {Rivas}}\ and\ \bibinfo {author} {\bibfnamefont {A.}~\bibnamefont {Luis}},\
  }\bibfield  {title} {\enquote {\bibinfo {title} {Sub-heisenberg estimation of
  non-random phase shifts},}\ }\href@noop {} {\bibfield  {journal} {\bibinfo
  {journal} {New J. Phys.}\ }\textbf {\bibinfo {volume} {14}},\ \bibinfo
  {pages} {093052} (\bibinfo {year} {2012})}\BibitemShut {NoStop}%
\bibitem [{\citenamefont {Braunstein}\ and\ \citenamefont
  {Caves}(1994)}]{Braunstein1994a}%
  \BibitemOpen
  \bibfield  {author} {\bibinfo {author} {\bibfnamefont {S.~L.}\ \bibnamefont
  {Braunstein}}\ and\ \bibinfo {author} {\bibfnamefont {C.~M.}\ \bibnamefont
  {Caves}},\ }\bibfield  {title} {\enquote {\bibinfo {title} {Statistical
  distance and the geometry of quantum states},}\ }\href@noop {} {\bibfield
  {journal} {\bibinfo  {journal} {Phys. Rev. Lett.}\ }\textbf {\bibinfo
  {volume} {72}},\ \bibinfo {pages} {3439} (\bibinfo {year}
  {1994})}\BibitemShut {NoStop}%
\bibitem [{\citenamefont {Braunstein}\ \emph {et~al.}()\citenamefont
  {Braunstein}, \citenamefont {Caves},\ and\ \citenamefont
  {Milburn}}]{Braunstein1996}%
  \BibitemOpen
  \bibfield  {author} {\bibinfo {author} {\bibfnamefont {S.~L.}\ \bibnamefont
  {Braunstein}}, \bibinfo {author} {\bibfnamefont {C.~M.}\ \bibnamefont
  {Caves}}, \ and\ \bibinfo {author} {\bibfnamefont {G.~J.}\ \bibnamefont
  {Milburn}},\ }\bibfield  {title} {\enquote {\bibinfo {title} {Generalized
  uncertainty relations: Theory, examples, and lorentz invariance},}\
  }\href@noop {} {\ }\BibitemShut {NoStop}%
\bibitem [{\citenamefont {Rubio}\ \emph {et~al.}(2018)\citenamefont {Rubio},
  \citenamefont {Knott},\ and\ \citenamefont {Dunningham}}]{Rubio2018}%
  \BibitemOpen
  \bibfield  {author} {\bibinfo {author} {\bibfnamefont {J.}~\bibnamefont
  {Rubio}}, \bibinfo {author} {\bibfnamefont {P.}~\bibnamefont {Knott}}, \ and\
  \bibinfo {author} {\bibfnamefont {J.}~\bibnamefont {Dunningham}},\ }\bibfield
   {title} {\enquote {\bibinfo {title} {Non-asymptotic analysis of quantum
  metrology protocols beyond the cram{\'e}r-rao bound},}\ }\href@noop {}
  {\bibfield  {journal} {\bibinfo  {journal} {J. Phys. Commun.}\ }\textbf
  {\bibinfo {volume} {2}},\ \bibinfo {pages} {015027} (\bibinfo {year}
  {2018})}\BibitemShut {NoStop}%
\bibitem [{\citenamefont {Rubio}\ and\ \citenamefont
  {Dunningham}()}]{Rubio2019}%
  \BibitemOpen
  \bibfield  {author} {\bibinfo {author} {\bibfnamefont {J.}~\bibnamefont
  {Rubio}}\ and\ \bibinfo {author} {\bibfnamefont {J.}~\bibnamefont
  {Dunningham}},\ }\bibfield  {title} {\enquote {\bibinfo {title} {Bayesian
  multi-parameter quantum metrology with limited data},}\ }\href@noop {}
  {\bibinfo  {journal} {arXiv:1906.04123}\ }\BibitemShut {NoStop}%
\bibitem [{\citenamefont {James}(2006)}]{James2006}%
  \BibitemOpen
\bibfield  {journal} {  }\bibfield  {author} {\bibinfo {author} {\bibfnamefont
  {F.}~\bibnamefont {James}},\ }\href@noop {} {\emph {\bibinfo {title}
  {Statistical Methods in Experimental Physics}}}\ (\bibinfo  {publisher}
  {World Scientific Publishing, Singapore},\ \bibinfo {year}
  {2006})\BibitemShut {NoStop}%
\bibitem [{\citenamefont {Bollinger}\ \emph {et~al.}(1996)\citenamefont
  {Bollinger}, \citenamefont {Itano}, \citenamefont {Wineland},\ and\
  \citenamefont {Heinzen}}]{Bollinger1996}%
  \BibitemOpen
  \bibfield  {author} {\bibinfo {author} {\bibfnamefont {J.~J.}\ \bibnamefont
  {Bollinger}}, \bibinfo {author} {\bibfnamefont {W.~M.}\ \bibnamefont
  {Itano}}, \bibinfo {author} {\bibfnamefont {D.~J.}\ \bibnamefont {Wineland}},
  \ and\ \bibinfo {author} {\bibfnamefont {D.~J.}\ \bibnamefont {Heinzen}},\
  }\bibfield  {title} {\enquote {\bibinfo {title} {Optimal frequency
  measurements with maximally correlated states},}\ }\href@noop {} {\bibfield
  {journal} {\bibinfo  {journal} {Phys. Rev. A}\ }\textbf {\bibinfo {volume}
  {54}},\ \bibinfo {pages} {R4649} (\bibinfo {year} {1996})}\BibitemShut
  {NoStop}%
\bibitem [{\citenamefont {Yurke}\ \emph {et~al.}(1986)\citenamefont {Yurke},
  \citenamefont {McCall},\ and\ \citenamefont {Klauder}}]{Yurke1986}%
  \BibitemOpen
  \bibfield  {author} {\bibinfo {author} {\bibfnamefont {B.}~\bibnamefont
  {Yurke}}, \bibinfo {author} {\bibfnamefont {S.~L.}\ \bibnamefont {McCall}}, \
  and\ \bibinfo {author} {\bibfnamefont {J.~R.}\ \bibnamefont {Klauder}},\
  }\bibfield  {title} {\enquote {\bibinfo {title} {\text{SU}(2) and
  \text{SU}(1,1) interferometers},}\ }\href@noop {} {\bibfield  {journal}
  {\bibinfo  {journal} {Phys. Rev. A}\ }\textbf {\bibinfo {volume} {33}},\
  \bibinfo {pages} {4033} (\bibinfo {year} {1986})}\BibitemShut {NoStop}%
\bibitem [{\citenamefont {Sanders}\ and\ \citenamefont
  {Milburn}(1995)}]{Sanders1995}%
  \BibitemOpen
  \bibfield  {author} {\bibinfo {author} {\bibfnamefont {B.~C.}\ \bibnamefont
  {Sanders}}\ and\ \bibinfo {author} {\bibfnamefont {G.~J.}\ \bibnamefont
  {Milburn}},\ }\bibfield  {title} {\enquote {\bibinfo {title} {Optimal quantum
  measurements for phase estimation},}\ }\href@noop {} {\bibfield  {journal}
  {\bibinfo  {journal} {Phys. Rev. Lett.}\ }\textbf {\bibinfo {volume} {75}},\
  \bibinfo {pages} {2944} (\bibinfo {year} {1995})}\BibitemShut {NoStop}%
\bibitem [{\citenamefont {Ou}(1996)}]{Ou1996}%
  \BibitemOpen
  \bibfield  {author} {\bibinfo {author} {\bibfnamefont {Z.~Y.}\ \bibnamefont
  {Ou}},\ }\bibfield  {title} {\enquote {\bibinfo {title} {Complementarity and
  fundamental limit in precision phase measurement},}\ }\href@noop {}
  {\bibfield  {journal} {\bibinfo  {journal} {Phys. Rev. Lett.}\ }\textbf
  {\bibinfo {volume} {77}},\ \bibinfo {pages} {2352} (\bibinfo {year}
  {1996})}\BibitemShut {NoStop}%
\bibitem [{\citenamefont {Zwierz}\ \emph {et~al.}(2010)\citenamefont {Zwierz},
  \citenamefont {P\'{e}rez-Delgado},\ and\ \citenamefont {Kok}}]{Zwierz2010}%
  \BibitemOpen
  \bibfield  {author} {\bibinfo {author} {\bibfnamefont {M.}~\bibnamefont
  {Zwierz}}, \bibinfo {author} {\bibfnamefont {C.~A.}\ \bibnamefont
  {P\'{e}rez-Delgado}}, \ and\ \bibinfo {author} {\bibfnamefont
  {P.}~\bibnamefont {Kok}},\ }\bibfield  {title} {\enquote {\bibinfo {title}
  {General optimality of the heisenberg limit for quantum metrology},}\
  }\href@noop {} {\bibfield  {journal} {\bibinfo  {journal} {Phys. Rev. Lett.}\
  }\textbf {\bibinfo {volume} {105}},\ \bibinfo {pages} {180402} (\bibinfo
  {year} {2010})}\BibitemShut {NoStop}%
\bibitem [{\citenamefont {Luis}\ and\ \citenamefont {Rodil}(2013)}]{Luis2013a}%
  \BibitemOpen
  \bibfield  {author} {\bibinfo {author} {\bibfnamefont {A.}~\bibnamefont
  {Luis}}\ and\ \bibinfo {author} {\bibfnamefont {A.}~\bibnamefont {Rodil}},\
  }\bibfield  {title} {\enquote {\bibinfo {title} {Alternative measures of
  uncertainty in quantum metrology: Contradictions and limits},}\ }\href@noop
  {} {\bibfield  {journal} {\bibinfo  {journal} {Phys. Rev. A}\ }\textbf
  {\bibinfo {volume} {87}},\ \bibinfo {pages} {034101} (\bibinfo {year}
  {2013})}\BibitemShut {NoStop}%
\bibitem [{\citenamefont {Luis}(2013)}]{Luis2013b}%
  \BibitemOpen
  \bibfield  {author} {\bibinfo {author} {\bibfnamefont {A.}~\bibnamefont
  {Luis}},\ }\bibfield  {title} {\enquote {\bibinfo {title} {Signal detection
  without finite-energy limits to quantum resolution},}\ }\href@noop {}
  {\bibfield  {journal} {\bibinfo  {journal} {Ann. Phys. (Amsterdam)}\ }\textbf
  {\bibinfo {volume} {331}},\ \bibinfo {pages} {1} (\bibinfo {year}
  {2013})}\BibitemShut {NoStop}%
\bibitem [{\citenamefont {Anisimov}\ \emph {et~al.}(2010)\citenamefont
  {Anisimov}, \citenamefont {Raterman}, \citenamefont {Chiruvelli},
  \citenamefont {Plick}, \citenamefont {Huver}, \citenamefont {Lee},\ and\
  \citenamefont {Dowling}}]{Anisimov2010}%
  \BibitemOpen
  \bibfield  {author} {\bibinfo {author} {\bibfnamefont {P.~M.}\ \bibnamefont
  {Anisimov}}, \bibinfo {author} {\bibfnamefont {G.~M.}\ \bibnamefont
  {Raterman}}, \bibinfo {author} {\bibfnamefont {A.}~\bibnamefont
  {Chiruvelli}}, \bibinfo {author} {\bibfnamefont {W.~N.}\ \bibnamefont
  {Plick}}, \bibinfo {author} {\bibfnamefont {S.~D.}\ \bibnamefont {Huver}},
  \bibinfo {author} {\bibfnamefont {H.}~\bibnamefont {Lee}}, \ and\ \bibinfo
  {author} {\bibfnamefont {J.~P.}\ \bibnamefont {Dowling}},\ }\bibfield
  {title} {\enquote {\bibinfo {title} {Quantum metrology with two-mode squeezed
  vacuum: Parity detection beats the heisenberg limit},}\ }\href@noop {}
  {\bibfield  {journal} {\bibinfo  {journal} {Phys. Rev. Lett.}\ }\textbf
  {\bibinfo {volume} {104}},\ \bibinfo {pages} {103602} (\bibinfo {year}
  {2010})}\BibitemShut {NoStop}%
\bibitem [{\citenamefont {Tsang}(2012)}]{Tsang2012}%
  \BibitemOpen
  \bibfield  {author} {\bibinfo {author} {\bibfnamefont {M.}~\bibnamefont
  {Tsang}},\ }\bibfield  {title} {\enquote {\bibinfo {title} {Ziv-\text{Z}akai
  error bounds for quantum parameter estimation},}\ }\href@noop {} {\bibfield
  {journal} {\bibinfo  {journal} {Phys. Rev. Lett.}\ }\textbf {\bibinfo
  {volume} {108}},\ \bibinfo {pages} {230401} (\bibinfo {year}
  {2012})}\BibitemShut {NoStop}%
\bibitem [{\citenamefont {Giovannetti}\ and\ \citenamefont
  {Maccone}(2012)}]{Giovannetti2012a}%
  \BibitemOpen
  \bibfield  {author} {\bibinfo {author} {\bibfnamefont {V.}~\bibnamefont
  {Giovannetti}}\ and\ \bibinfo {author} {\bibfnamefont {L.}~\bibnamefont
  {Maccone}},\ }\bibfield  {title} {\enquote {\bibinfo {title} {Sub-heisenberg
  estimation strategies are ineffective},}\ }\href@noop {} {\bibfield
  {journal} {\bibinfo  {journal} {Phys. Rev. Lett.}\ }\textbf {\bibinfo
  {volume} {108}},\ \bibinfo {pages} {210404} (\bibinfo {year}
  {2012})}\BibitemShut {NoStop}%
\bibitem [{\citenamefont {Giovannetti}\ \emph {et~al.}(2012)\citenamefont
  {Giovannetti}, \citenamefont {Lloyd},\ and\ \citenamefont
  {Maccone}}]{Giovannetti2012b}%
  \BibitemOpen
  \bibfield  {author} {\bibinfo {author} {\bibfnamefont {V.}~\bibnamefont
  {Giovannetti}}, \bibinfo {author} {\bibfnamefont {S.}~\bibnamefont {Lloyd}},
  \ and\ \bibinfo {author} {\bibfnamefont {L.}~\bibnamefont {Maccone}},\
  }\bibfield  {title} {\enquote {\bibinfo {title} {Quantum measurement bounds
  beyond the uncertainty relations},}\ }\href@noop {} {\bibfield  {journal}
  {\bibinfo  {journal} {Phys. Rev. Lett.}\ }\textbf {\bibinfo {volume} {108}},\
  \bibinfo {pages} {260405} (\bibinfo {year} {2012})}\BibitemShut {NoStop}%
\bibitem [{\citenamefont {Berry}\ \emph {et~al.}(2012)\citenamefont {Berry},
  \citenamefont {Hall}, \citenamefont {Zwierz},\ and\ \citenamefont
  {Wiseman}}]{Berry2012}%
  \BibitemOpen
  \bibfield  {author} {\bibinfo {author} {\bibfnamefont {D.~W.}\ \bibnamefont
  {Berry}}, \bibinfo {author} {\bibfnamefont {M.~J.~W.}\ \bibnamefont {Hall}},
  \bibinfo {author} {\bibfnamefont {M.}~\bibnamefont {Zwierz}}, \ and\ \bibinfo
  {author} {\bibfnamefont {H.~M.}\ \bibnamefont {Wiseman}},\ }\bibfield
  {title} {\enquote {\bibinfo {title} {Optimal heisenberg-style bounds for the
  average performance of arbitrary phase estimates},}\ }\href@noop {}
  {\bibfield  {journal} {\bibinfo  {journal} {Phys. Rev. A}\ }\textbf {\bibinfo
  {volume} {86}},\ \bibinfo {pages} {053813} (\bibinfo {year}
  {2012})}\BibitemShut {NoStop}%
\bibitem [{\citenamefont {Hall}\ \emph {et~al.}(2012)\citenamefont {Hall},
  \citenamefont {Berry}, \citenamefont {Zwierz},\ and\ \citenamefont
  {Wiseman}}]{Hall2012a}%
  \BibitemOpen
  \bibfield  {author} {\bibinfo {author} {\bibfnamefont {M.~J.~W.}\
  \bibnamefont {Hall}}, \bibinfo {author} {\bibfnamefont {D.~W.}\ \bibnamefont
  {Berry}}, \bibinfo {author} {\bibfnamefont {M.}~\bibnamefont {Zwierz}}, \
  and\ \bibinfo {author} {\bibfnamefont {H.~M.}\ \bibnamefont {Wiseman}},\
  }\bibfield  {title} {\enquote {\bibinfo {title} {Universality of the
  heisenberg limit for estimates of random phase shifts},}\ }\href@noop {}
  {\bibfield  {journal} {\bibinfo  {journal} {Phys. Rev. A}\ }\textbf {\bibinfo
  {volume} {85}},\ \bibinfo {pages} {041802} (\bibinfo {year}
  {2012})}\BibitemShut {NoStop}%
\bibitem [{\citenamefont {Hall}\ and\ \citenamefont
  {Wiseman}(2012{\natexlab{a}})}]{Hall2012b}%
  \BibitemOpen
  \bibfield  {author} {\bibinfo {author} {\bibfnamefont {M.~J.~W.}\
  \bibnamefont {Hall}}\ and\ \bibinfo {author} {\bibfnamefont {H.~M.}\
  \bibnamefont {Wiseman}},\ }\bibfield  {title} {\enquote {\bibinfo {title}
  {Heisenberg-style bounds for arbitrary estimates of shift parameters
  including prior information},}\ }\href@noop {} {\bibfield  {journal}
  {\bibinfo  {journal} {New J. Phys.}\ }\textbf {\bibinfo {volume} {14}},\
  \bibinfo {pages} {033040} (\bibinfo {year} {2012}{\natexlab{a}})}\BibitemShut
  {NoStop}%
\bibitem [{\citenamefont {Hall}\ and\ \citenamefont
  {Wiseman}(2012{\natexlab{b}})}]{Hall2012c}%
  \BibitemOpen
  \bibfield  {author} {\bibinfo {author} {\bibfnamefont {M.~J.~W.}\
  \bibnamefont {Hall}}\ and\ \bibinfo {author} {\bibfnamefont {H.~M.}\
  \bibnamefont {Wiseman}},\ }\bibfield  {title} {\enquote {\bibinfo {title}
  {Does nonlinear metrology offer improved resolution?}}\ }\href@noop {}
  {\bibfield  {journal} {\bibinfo  {journal} {Phys. Rev. X}\ }\textbf {\bibinfo
  {volume} {2}},\ \bibinfo {pages} {041006} (\bibinfo {year}
  {2012}{\natexlab{b}})}\BibitemShut {NoStop}%
\bibitem [{\citenamefont {Jarzyna}\ and\ \citenamefont {Demkowicz-Dobrza{\'
  n}ski}(2015)}]{Jarzyna2015}%
  \BibitemOpen
  \bibfield  {author} {\bibinfo {author} {\bibfnamefont {M.}~\bibnamefont
  {Jarzyna}}\ and\ \bibinfo {author} {\bibfnamefont {R.}~\bibnamefont
  {Demkowicz-Dobrza{\' n}ski}},\ }\bibfield  {title} {\enquote {\bibinfo
  {title} {True precision limits in quantum metrology},}\ }\href@noop {}
  {\bibfield  {journal} {\bibinfo  {journal} {New J. Phys.}\ }\textbf {\bibinfo
  {volume} {17}},\ \bibinfo {pages} {013010} (\bibinfo {year}
  {2015})}\BibitemShut {NoStop}%
\bibitem [{\citenamefont {Lee}\ \emph {et~al.}(2017)\citenamefont {Lee},
  \citenamefont {Huynh}, \citenamefont {Lee}, \citenamefont {Lee},
  \citenamefont {Lee}, \citenamefont {Tame}, \citenamefont {Rockstuhl},\ and\
  \citenamefont {Lee}}]{Lee2017}%
  \BibitemOpen
  \bibfield  {author} {\bibinfo {author} {\bibfnamefont {J.-S.}\ \bibnamefont
  {Lee}}, \bibinfo {author} {\bibfnamefont {T.}~\bibnamefont {Huynh}}, \bibinfo
  {author} {\bibfnamefont {S.-Y.}\ \bibnamefont {Lee}}, \bibinfo {author}
  {\bibfnamefont {K.-G.}\ \bibnamefont {Lee}}, \bibinfo {author} {\bibfnamefont
  {J.}~\bibnamefont {Lee}}, \bibinfo {author} {\bibfnamefont {M.}~\bibnamefont
  {Tame}}, \bibinfo {author} {\bibfnamefont {C.}~\bibnamefont {Rockstuhl}}, \
  and\ \bibinfo {author} {\bibfnamefont {C.}~\bibnamefont {Lee}},\ }\bibfield
  {title} {\enquote {\bibinfo {title} {Quantum noise reduction in
  intensity-sensitive surface-plasmon-resonance sensors},}\ }\href@noop {}
  {\bibfield  {journal} {\bibinfo  {journal} {Phys. Rev. A}\ }\textbf {\bibinfo
  {volume} {96}},\ \bibinfo {pages} {033833} (\bibinfo {year}
  {2017})}\BibitemShut {NoStop}%
\bibitem [{\citenamefont {Lee}\ \emph {et~al.}(2018)\citenamefont {Lee},
  \citenamefont {Yoon}, \citenamefont {Rah}, \citenamefont {Tame},
  \citenamefont {Rockstuhl}, \citenamefont {Song}, \citenamefont {Lee},\ and\
  \citenamefont {Lee}}]{Lee2018}%
  \BibitemOpen
  \bibfield  {author} {\bibinfo {author} {\bibfnamefont {J.-S.}\ \bibnamefont
  {Lee}}, \bibinfo {author} {\bibfnamefont {S.-J.}\ \bibnamefont {Yoon}},
  \bibinfo {author} {\bibfnamefont {H.}~\bibnamefont {Rah}}, \bibinfo {author}
  {\bibfnamefont {M.}~\bibnamefont {Tame}}, \bibinfo {author} {\bibfnamefont
  {C.}~\bibnamefont {Rockstuhl}}, \bibinfo {author} {\bibfnamefont {S.~H.}\
  \bibnamefont {Song}}, \bibinfo {author} {\bibfnamefont {C.}~\bibnamefont
  {Lee}}, \ and\ \bibinfo {author} {\bibfnamefont {K.-G.}\ \bibnamefont
  {Lee}},\ }\bibfield  {title} {\enquote {\bibinfo {title} {Quantum plasmonic
  sensing using single photons},}\ }\href@noop {} {\bibfield  {journal}
  {\bibinfo  {journal} {Opt. Express}\ }\textbf {\bibinfo {volume} {26}},\
  \bibinfo {pages} {29272} (\bibinfo {year} {2018})}\BibitemShut {NoStop}%
\bibitem [{\citenamefont {Yonezawa}\ \emph {et~al.}(2012)\citenamefont
  {Yonezawa}, \citenamefont {Nakane}, \citenamefont {Wheatley}, \citenamefont
  {Iwasawa}, \citenamefont {Takeda}, \citenamefont {Arao}, \citenamefont
  {Ohki}, \citenamefont {Tsumura}, \citenamefont {Berry}, \citenamefont
  {Ralph}, \citenamefont {Wiseman}, \citenamefont {Huntington},\ and\
  \citenamefont {Furusawa}}]{Yonezawa2012}%
  \BibitemOpen
  \bibfield  {author} {\bibinfo {author} {\bibfnamefont {H.}~\bibnamefont
  {Yonezawa}}, \bibinfo {author} {\bibfnamefont {D.}~\bibnamefont {Nakane}},
  \bibinfo {author} {\bibfnamefont {T.~A.}\ \bibnamefont {Wheatley}}, \bibinfo
  {author} {\bibfnamefont {K.}~\bibnamefont {Iwasawa}}, \bibinfo {author}
  {\bibfnamefont {S.}~\bibnamefont {Takeda}}, \bibinfo {author} {\bibfnamefont
  {H.}~\bibnamefont {Arao}}, \bibinfo {author} {\bibfnamefont {K.}~\bibnamefont
  {Ohki}}, \bibinfo {author} {\bibfnamefont {K.}~\bibnamefont {Tsumura}},
  \bibinfo {author} {\bibfnamefont {D.~W.}\ \bibnamefont {Berry}}, \bibinfo
  {author} {\bibfnamefont {T.~C.}\ \bibnamefont {Ralph}}, \bibinfo {author}
  {\bibfnamefont {H.~M.}\ \bibnamefont {Wiseman}}, \bibinfo {author}
  {\bibfnamefont {E.~H.}\ \bibnamefont {Huntington}}, \ and\ \bibinfo {author}
  {\bibfnamefont {A.}~\bibnamefont {Furusawa}},\ }\bibfield  {title} {\enquote
  {\bibinfo {title} {Quantum-enhanced optical-phase tracking},}\ }\href@noop {}
  {\bibfield  {journal} {\bibinfo  {journal} {Science}\ }\textbf {\bibinfo
  {volume} {337}},\ \bibinfo {pages} {1514} (\bibinfo {year}
  {2012})}\BibitemShut {NoStop}%
\bibitem [{\citenamefont {McCormick}\ \emph {et~al.}(2019)\citenamefont
  {McCormick}, \citenamefont {Keller}, \citenamefont {Burd}, \citenamefont
  {Wineland}, \citenamefont {Wilson},\ and\ \citenamefont
  {Leibfried}}]{McCormick2019}%
  \BibitemOpen
  \bibfield  {author} {\bibinfo {author} {\bibfnamefont {K.~C.}\ \bibnamefont
  {McCormick}}, \bibinfo {author} {\bibfnamefont {J.}~\bibnamefont {Keller}},
  \bibinfo {author} {\bibfnamefont {S.~C.}\ \bibnamefont {Burd}}, \bibinfo
  {author} {\bibfnamefont {D.~J.}\ \bibnamefont {Wineland}}, \bibinfo {author}
  {\bibfnamefont {A.~C.}\ \bibnamefont {Wilson}}, \ and\ \bibinfo {author}
  {\bibfnamefont {D.}~\bibnamefont {Leibfried}},\ }\bibfield  {title} {\enquote
  {\bibinfo {title} {Quantum-enhanced sensing of a single-ion mechanical
  oscillator},}\ }\href@noop {} {\bibfield  {journal} {\bibinfo  {journal}
  {Nature}\ }\textbf {\bibinfo {volume} {572}},\ \bibinfo {pages} {86}
  (\bibinfo {year} {2019})}\BibitemShut {NoStop}%
\bibitem [{\citenamefont {F.~Dell’Anno}\ and\ \citenamefont
  {Illuminati}(2006)}]{DellAnno2006}%
  \BibitemOpen
  \bibfield  {author} {\bibinfo {author} {\bibfnamefont {S.~De~Siena}\
  \bibnamefont {F.~Dell’Anno}}\ and\ \bibinfo {author} {\bibfnamefont
  {F.}~\bibnamefont {Illuminati}},\ }\bibfield  {title} {\enquote {\bibinfo
  {title} {Multiphoton quantum optics and quantum state engineering},}\
  }\href@noop {} {\bibfield  {journal} {\bibinfo  {journal} {Phys. Rep.}\
  }\textbf {\bibinfo {volume} {428}},\ \bibinfo {pages} {53} (\bibinfo {year}
  {2006})}\BibitemShut {NoStop}%
\bibitem [{\citenamefont {Bimbard}\ \emph {et~al.}(2010)\citenamefont
  {Bimbard}, \citenamefont {Jain}, \citenamefont {MacRae},\ and\ \citenamefont
  {Lvovsky}}]{Bimbard2010}%
  \BibitemOpen
  \bibfield  {author} {\bibinfo {author} {\bibfnamefont {E.}~\bibnamefont
  {Bimbard}}, \bibinfo {author} {\bibfnamefont {N.}~\bibnamefont {Jain}},
  \bibinfo {author} {\bibfnamefont {A.}~\bibnamefont {MacRae}}, \ and\ \bibinfo
  {author} {\bibfnamefont {A.~I.}\ \bibnamefont {Lvovsky}},\ }\bibfield
  {title} {\enquote {\bibinfo {title} {Quantum-optical state engineering up to
  the two-photon level},}\ }\href@noop {} {\bibfield  {journal} {\bibinfo
  {journal} {Nat. Photon.}\ }\textbf {\bibinfo {volume} {4}},\ \bibinfo {pages}
  {243} (\bibinfo {year} {2010})}\BibitemShut {NoStop}%
\bibitem [{\citenamefont {Nichols}\ \emph {et~al.}()\citenamefont {Nichols},
  \citenamefont {Mineh}, \citenamefont {Rubio}, \citenamefont {Matthews},\ and\
  \citenamefont {Knott}}]{Nichols2018}%
  \BibitemOpen
  \bibfield  {author} {\bibinfo {author} {\bibfnamefont {R.}~\bibnamefont
  {Nichols}}, \bibinfo {author} {\bibfnamefont {L.}~\bibnamefont {Mineh}},
  \bibinfo {author} {\bibfnamefont {J.}~\bibnamefont {Rubio}}, \bibinfo
  {author} {\bibfnamefont {J.~C.~F.}\ \bibnamefont {Matthews}}, \ and\ \bibinfo
  {author} {\bibfnamefont {P.~A.}\ \bibnamefont {Knott}},\ }\bibfield  {title}
  {\enquote {\bibinfo {title} {Designing quantum experiments with a genetic
  algorithm},}\ }\href@noop {} {\bibinfo  {journal} {arXiv:1812.01032}\
  }\BibitemShut {NoStop}%
\bibitem [{\citenamefont {Arrazola}\ \emph {et~al.}(2019)\citenamefont
  {Arrazola}, \citenamefont {Bromley}, \citenamefont {Izaac}, \citenamefont
  {Myers}, \citenamefont {Br{\'a}dler},\ and\ \citenamefont
  {Killoran}}]{Arrazola2019}%
  \BibitemOpen
\bibfield  {journal} {  }\bibfield  {author} {\bibinfo {author} {\bibfnamefont
  {J.~M.}\ \bibnamefont {Arrazola}}, \bibinfo {author} {\bibfnamefont {T.~R.}\
  \bibnamefont {Bromley}}, \bibinfo {author} {\bibfnamefont {J.}~\bibnamefont
  {Izaac}}, \bibinfo {author} {\bibfnamefont {C.~R.}\ \bibnamefont {Myers}},
  \bibinfo {author} {\bibfnamefont {K.}~\bibnamefont {Br{\'a}dler}}, \ and\
  \bibinfo {author} {\bibfnamefont {N.}~\bibnamefont {Killoran}},\ }\bibfield
  {title} {\enquote {\bibinfo {title} {Machine learning method for state
  preparation and gate synthesis on photonic quantum computers},}\ }\href@noop
  {} {\bibfield  {journal} {\bibinfo  {journal} {Quantum Sci. Technol.}\
  }\textbf {\bibinfo {volume} {4}},\ \bibinfo {pages} {024004} (\bibinfo {year}
  {2019})}\BibitemShut {NoStop}%
\bibitem [{\citenamefont {O’Driscoll}\ \emph {et~al.}(2019)\citenamefont
  {O’Driscoll}, \citenamefont {Nichols},\ and\ \citenamefont
  {Knott}}]{ODriscoll2019}%
  \BibitemOpen
  \bibfield  {author} {\bibinfo {author} {\bibfnamefont {L.}~\bibnamefont
  {O’Driscoll}}, \bibinfo {author} {\bibfnamefont {R.}~\bibnamefont
  {Nichols}}, \ and\ \bibinfo {author} {\bibfnamefont {P.~A.}\ \bibnamefont
  {Knott}},\ }\bibfield  {title} {\enquote {\bibinfo {title} {A hybrid machine
  learning algorithm for designing quantum experiments},}\ }\href@noop {}
  {\bibfield  {journal} {\bibinfo  {journal} {Quantum Mach. Intell.}\ }\textbf
  {\bibinfo {volume} {1}},\ \bibinfo {pages} {5} (\bibinfo {year}
  {2019})}\BibitemShut {NoStop}%
\bibitem [{\citenamefont {Hofmann}(2009)}]{Hofmann2009}%
  \BibitemOpen
  \bibfield  {author} {\bibinfo {author} {\bibfnamefont {H.~F.}\ \bibnamefont
  {Hofmann}},\ }\bibfield  {title} {\enquote {\bibinfo {title} {All
  path-symmetric pure states achieve their maximal phase sensitivity in
  conventional two-path interferometry},}\ }\href@noop {} {\bibfield  {journal}
  {\bibinfo  {journal} {Phys. Rev. A}\ }\textbf {\bibinfo {volume} {79}},\
  \bibinfo {pages} {033822} (\bibinfo {year} {2009})}\BibitemShut {NoStop}%
\bibitem [{\citenamefont {Hyllus}\ \emph {et~al.}(2010)\citenamefont {Hyllus},
  \citenamefont {Pezz{\'e}},\ and\ \citenamefont {Smerzi}}]{Hyllus2010}%
  \BibitemOpen
  \bibfield  {author} {\bibinfo {author} {\bibfnamefont {P.}~\bibnamefont
  {Hyllus}}, \bibinfo {author} {\bibfnamefont {L.}~\bibnamefont {Pezz{\'e}}}, \
  and\ \bibinfo {author} {\bibfnamefont {A.}~\bibnamefont {Smerzi}},\
  }\bibfield  {title} {\enquote {\bibinfo {title} {Entanglement and sensitivity
  in precision measurements with states of a fluctuating number of
  particles},}\ }\href@noop {} {\bibfield  {journal} {\bibinfo  {journal}
  {Phys. Rev. Lett.}\ }\textbf {\bibinfo {volume} {105}},\ \bibinfo {pages}
  {120501} (\bibinfo {year} {2010})}\BibitemShut {NoStop}%
\bibitem [{\citenamefont {Leonhardt}\ \emph {et~al.}(1995)\citenamefont
  {Leonhardt}, \citenamefont {Vaccaro}, \citenamefont {B{\" o}hmer},\ and\
  \citenamefont {Paul}}]{Leonhardt1995}%
  \BibitemOpen
  \bibfield  {author} {\bibinfo {author} {\bibfnamefont {U.}~\bibnamefont
  {Leonhardt}}, \bibinfo {author} {\bibfnamefont {J.~A.}\ \bibnamefont
  {Vaccaro}}, \bibinfo {author} {\bibfnamefont {B.}~\bibnamefont {B{\"
  o}hmer}}, \ and\ \bibinfo {author} {\bibfnamefont {H.}~\bibnamefont {Paul}},\
  }\bibfield  {title} {\enquote {\bibinfo {title} {Canonical and measured phase
  distributions},}\ }\href@noop {} {\bibfield  {journal} {\bibinfo  {journal}
  {Phys. Rev. A}\ }\textbf {\bibinfo {volume} {51}},\ \bibinfo {pages} {84}
  (\bibinfo {year} {1995})}\BibitemShut {NoStop}%
\bibitem [{\citenamefont {Pirandola}\ \emph {et~al.}(2018)\citenamefont
  {Pirandola}, \citenamefont {Bardhan}, \citenamefont {Gehring}, \citenamefont
  {Weedbrook},\ and\ \citenamefont {Lloyd}}]{Pirandola2018}%
  \BibitemOpen
  \bibfield  {author} {\bibinfo {author} {\bibfnamefont {S.}~\bibnamefont
  {Pirandola}}, \bibinfo {author} {\bibfnamefont {B.~R.}\ \bibnamefont
  {Bardhan}}, \bibinfo {author} {\bibfnamefont {T.}~\bibnamefont {Gehring}},
  \bibinfo {author} {\bibfnamefont {C.}~\bibnamefont {Weedbrook}}, \ and\
  \bibinfo {author} {\bibfnamefont {S.}~\bibnamefont {Lloyd}},\ }\bibfield
  {title} {\enquote {\bibinfo {title} {Advances in photonic quantum sensing},}\
  }\href@noop {} {\bibfield  {journal} {\bibinfo  {journal} {Nat. Photon.}\
  }\textbf {\bibinfo {volume} {12}},\ \bibinfo {pages} {724} (\bibinfo {year}
  {2018})}\BibitemShut {NoStop}%
\bibitem [{\citenamefont {Vahlbruch}\ \emph {et~al.}(2016)\citenamefont
  {Vahlbruch}, \citenamefont {Mehmet}, \citenamefont {Danzmann},\ and\
  \citenamefont {Schnabel}}]{Vahlbruch2016}%
  \BibitemOpen
  \bibfield  {author} {\bibinfo {author} {\bibfnamefont {H.}~\bibnamefont
  {Vahlbruch}}, \bibinfo {author} {\bibfnamefont {M.}~\bibnamefont {Mehmet}},
  \bibinfo {author} {\bibfnamefont {K.}~\bibnamefont {Danzmann}}, \ and\
  \bibinfo {author} {\bibfnamefont {R.}~\bibnamefont {Schnabel}},\ }\bibfield
  {title} {\enquote {\bibinfo {title} {Detection of 15 db squeezed states of
  light and their application for the absolute calibration of photoelectric
  quantum efficiency},}\ }\href@noop {} {\bibfield  {journal} {\bibinfo
  {journal} {Phys. Rev. Lett.}\ }\textbf {\bibinfo {volume} {117}},\ \bibinfo
  {pages} {110801} (\bibinfo {year} {2016})}\BibitemShut {NoStop}%
\bibitem [{\citenamefont {Aspachs}\ \emph {et~al.}(2009)\citenamefont
  {Aspachs}, \citenamefont {Calsamiglia}, \citenamefont {Mu{\~{n}}oz-Tapia},\
  and\ \citenamefont {Bagan}}]{Aspachs2009}%
  \BibitemOpen
  \bibfield  {author} {\bibinfo {author} {\bibfnamefont {M.}~\bibnamefont
  {Aspachs}}, \bibinfo {author} {\bibfnamefont {J.}~\bibnamefont
  {Calsamiglia}}, \bibinfo {author} {\bibfnamefont {R.}~\bibnamefont
  {Mu{\~{n}}oz-Tapia}}, \ and\ \bibinfo {author} {\bibfnamefont
  {E.}~\bibnamefont {Bagan}},\ }\bibfield  {title} {\enquote {\bibinfo {title}
  {Phase estimation for thermal gaussian states},}\ }\href@noop {} {\bibfield
  {journal} {\bibinfo  {journal} {Phys. Rev. A}\ }\textbf {\bibinfo {volume}
  {79}},\ \bibinfo {pages} {033834} (\bibinfo {year} {2009})}\BibitemShut
  {NoStop}%
\bibitem [{\citenamefont {{\v S}afr{\'a}nek}\ and\ \citenamefont
  {Fuentes}(2016)}]{Safranek2016}%
  \BibitemOpen
  \bibfield  {author} {\bibinfo {author} {\bibfnamefont {D.}~\bibnamefont {{\v
  S}afr{\'a}nek}}\ and\ \bibinfo {author} {\bibfnamefont {I.}~\bibnamefont
  {Fuentes}},\ }\bibfield  {title} {\enquote {\bibinfo {title} {Optimal probe
  states for the estimation of gaussian unitary channels},}\ }\href@noop {}
  {\bibfield  {journal} {\bibinfo  {journal} {Phys. Rev. A}\ }\textbf {\bibinfo
  {volume} {94}},\ \bibinfo {pages} {062313} (\bibinfo {year}
  {2016})}\BibitemShut {NoStop}%
\bibitem [{\citenamefont {Oh}\ \emph {et~al.}()\citenamefont {Oh},
  \citenamefont {Lee}, \citenamefont {Banchi}, \citenamefont {Lee},
  \citenamefont {Rockstuhl},\ and\ \citenamefont {Jeong}}]{Oh2019b}%
  \BibitemOpen
  \bibfield  {author} {\bibinfo {author} {\bibfnamefont {C.}~\bibnamefont
  {Oh}}, \bibinfo {author} {\bibfnamefont {C.}~\bibnamefont {Lee}}, \bibinfo
  {author} {\bibfnamefont {L.}~\bibnamefont {Banchi}}, \bibinfo {author}
  {\bibfnamefont {S.-Y.}\ \bibnamefont {Lee}}, \bibinfo {author} {\bibfnamefont
  {C.}~\bibnamefont {Rockstuhl}}, \ and\ \bibinfo {author} {\bibfnamefont
  {H.}~\bibnamefont {Jeong}},\ }\bibfield  {title} {\enquote {\bibinfo {title}
  {Optimal measurements for quantum fidelity between gaussian states and its
  relevance to quantum metrology},}\ }\href@noop {} {\bibinfo  {journal} {Phys.
  Rev. A}\ }\BibitemShut {NoStop}%
\bibitem [{\citenamefont {Popoviciu}(1935)}]{Popoviciu1935}%
  \BibitemOpen
\bibfield  {journal} {  }\bibfield  {author} {\bibinfo {author} {\bibfnamefont
  {T.}~\bibnamefont {Popoviciu}},\ }\bibfield  {title} {\enquote {\bibinfo
  {title} {Sur les \'{e}quations alg\'{e}briques ayant toutes leurs racines
  r\'{e}elles},}\ }\href@noop {} {\bibfield  {journal} {\bibinfo  {journal}
  {Mathematica (Cluj)}\ }\textbf {\bibinfo {volume} {9}},\ \bibinfo {pages}
  {129} (\bibinfo {year} {1935})}\BibitemShut {NoStop}%
\bibitem [{\citenamefont {Bhatia}\ and\ \citenamefont
  {Davis}(2000)}]{Bhatia-Davis2000}%
  \BibitemOpen
  \bibfield  {author} {\bibinfo {author} {\bibfnamefont {R.}~\bibnamefont
  {Bhatia}}\ and\ \bibinfo {author} {\bibfnamefont {C.}~\bibnamefont {Davis}},\
  }\bibfield  {title} {\enquote {\bibinfo {title} {A better bound on the
  variance},}\ }\href@noop {} {\bibfield  {journal} {\bibinfo  {journal} {Am.
  Math. Mon.}\ }\textbf {\bibinfo {volume} {107}},\ \bibinfo {pages} {353}
  (\bibinfo {year} {2000})}\BibitemShut {NoStop}%
\bibitem [{\citenamefont {Sabapathy}\ and\ \citenamefont
  {Weedbrook}(2018)}]{Sabapathy2018}%
  \BibitemOpen
  \bibfield  {author} {\bibinfo {author} {\bibfnamefont {K.~K.}\ \bibnamefont
  {Sabapathy}}\ and\ \bibinfo {author} {\bibfnamefont {C.}~\bibnamefont
  {Weedbrook}},\ }\bibfield  {title} {\enquote {\bibinfo {title} {On states as
  resource units for universal quantum computation with photonic
  architectures},}\ }\href@noop {} {\bibfield  {journal} {\bibinfo  {journal}
  {Phys. Rev. A}\ }\textbf {\bibinfo {volume} {97}},\ \bibinfo {pages} {062315}
  (\bibinfo {year} {2018})}\BibitemShut {NoStop}%
\bibitem [{\citenamefont {Yukawa}\ \emph {et~al.}(2013)\citenamefont {Yukawa},
  \citenamefont {Miyata}, \citenamefont {Mizuta}, \citenamefont {Yonezawa},
  \citenamefont {Marek}, \citenamefont {Filip},\ and\ \citenamefont
  {Furusawa}}]{Yukawa2013}%
  \BibitemOpen
  \bibfield  {author} {\bibinfo {author} {\bibfnamefont {M.}~\bibnamefont
  {Yukawa}}, \bibinfo {author} {\bibfnamefont {K.}~\bibnamefont {Miyata}},
  \bibinfo {author} {\bibfnamefont {T.}~\bibnamefont {Mizuta}}, \bibinfo
  {author} {\bibfnamefont {H.}~\bibnamefont {Yonezawa}}, \bibinfo {author}
  {\bibfnamefont {P.}~\bibnamefont {Marek}}, \bibinfo {author} {\bibfnamefont
  {R.}~\bibnamefont {Filip}}, \ and\ \bibinfo {author} {\bibfnamefont
  {A.}~\bibnamefont {Furusawa}},\ }\bibfield  {title} {\enquote {\bibinfo
  {title} {Generating superposition of up-to three photons for continuous
  variable quantum information processing},}\ }\href@noop {} {\bibfield
  {journal} {\bibinfo  {journal} {Opt. Express}\ }\textbf {\bibinfo {volume}
  {21}},\ \bibinfo {pages} {5529} (\bibinfo {year} {2013})}\BibitemShut
  {NoStop}%
\bibitem [{\citenamefont {Luis}(2017)}]{Luis2017}%
  \BibitemOpen
  \bibfield  {author} {\bibinfo {author} {\bibfnamefont {A.}~\bibnamefont
  {Luis}},\ }\bibfield  {title} {\enquote {\bibinfo {title} {Breaking the weak
  heisenberg limit},}\ }\href@noop {} {\bibfield  {journal} {\bibinfo
  {journal} {Phys. Rev. A}\ }\textbf {\bibinfo {volume} {95}},\ \bibinfo
  {pages} {032113} (\bibinfo {year} {2017})}\BibitemShut {NoStop}%
\bibitem [{\citenamefont {Johnson}\ \emph {et~al.}(2005)\citenamefont
  {Johnson}, \citenamefont {Kemp},\ and\ \citenamefont {Kotz}}]{Johnson2005}%
  \BibitemOpen
  \bibfield  {author} {\bibinfo {author} {\bibfnamefont {N.~L.}\ \bibnamefont
  {Johnson}}, \bibinfo {author} {\bibfnamefont {A.~W.}\ \bibnamefont {Kemp}}, \
  and\ \bibinfo {author} {\bibfnamefont {S.}~\bibnamefont {Kotz}},\ }\href@noop
  {} {\emph {\bibinfo {title} {Univariate discrete distributions (3 ed.)}}}\
  (\bibinfo  {publisher} {John Wiley \& Sons.},\ \bibinfo {year}
  {2005})\BibitemShut {NoStop}%
\bibitem [{Note1()}]{Note1}%
  \BibitemOpen
  \bibinfo {note} {In the literature, the terms ``super-Heisenberg scaling''
  and ``sub-Heisenberg scaling'' have interchangeably used to denote the same
  limit~\cite {Boixo2008, Woolley2008, Rams2018, Roy2019}.}\BibitemShut {Stop}%
\bibitem [{\citenamefont {Boixo}\ \emph {et~al.}(2007)\citenamefont {Boixo},
  \citenamefont {Flammia}, \citenamefont {Caves},\ and\ \citenamefont
  {Geremia}}]{Boixo2007}%
  \BibitemOpen
  \bibfield  {author} {\bibinfo {author} {\bibfnamefont {S.}~\bibnamefont
  {Boixo}}, \bibinfo {author} {\bibfnamefont {S.~T.}\ \bibnamefont {Flammia}},
  \bibinfo {author} {\bibfnamefont {C.~M.}\ \bibnamefont {Caves}}, \ and\
  \bibinfo {author} {\bibfnamefont {JM}~\bibnamefont {Geremia}},\ }\bibfield
  {title} {\enquote {\bibinfo {title} {Generalized limits for single-parameter
  quantum estimation},}\ }\href@noop {} {\bibfield  {journal} {\bibinfo
  {journal} {Phys. Rev. Lett.}\ }\textbf {\bibinfo {volume} {98}},\ \bibinfo
  {pages} {090401} (\bibinfo {year} {2007})}\BibitemShut {NoStop}%
\bibitem [{\citenamefont {Boixo}\ \emph {et~al.}(2008)\citenamefont {Boixo},
  \citenamefont {Datta}, \citenamefont {Davis}, \citenamefont {Flammia},
  \citenamefont {Shaji},\ and\ \citenamefont {Caves}}]{Boixo2008}%
  \BibitemOpen
  \bibfield  {author} {\bibinfo {author} {\bibfnamefont {S.}~\bibnamefont
  {Boixo}}, \bibinfo {author} {\bibfnamefont {A.}~\bibnamefont {Datta}},
  \bibinfo {author} {\bibfnamefont {M.~J.}\ \bibnamefont {Davis}}, \bibinfo
  {author} {\bibfnamefont {S.~T.}\ \bibnamefont {Flammia}}, \bibinfo {author}
  {\bibfnamefont {A.}~\bibnamefont {Shaji}}, \ and\ \bibinfo {author}
  {\bibfnamefont {C.~M.}\ \bibnamefont {Caves}},\ }\bibfield  {title} {\enquote
  {\bibinfo {title} {Quantum metrology: Dynamics versus entanglement},}\
  }\href@noop {} {\bibfield  {journal} {\bibinfo  {journal} {Phys. Rev. Lett.}\
  }\textbf {\bibinfo {volume} {101}},\ \bibinfo {pages} {040403} (\bibinfo
  {year} {2008})}\BibitemShut {NoStop}%
\bibitem [{\citenamefont {Choi}\ and\ \citenamefont
  {Sundaram}(2008)}]{Choi2008}%
  \BibitemOpen
  \bibfield  {author} {\bibinfo {author} {\bibfnamefont {S.}~\bibnamefont
  {Choi}}\ and\ \bibinfo {author} {\bibfnamefont {B.}~\bibnamefont
  {Sundaram}},\ }\bibfield  {title} {\enquote {\bibinfo {title} {Bose-einstein
  condensate as a nonlinear ramsey interferometer operating beyond the
  heisenberg limit},}\ }\href@noop {} {\bibfield  {journal} {\bibinfo
  {journal} {Phys. Rev. A}\ }\textbf {\bibinfo {volume} {77}},\ \bibinfo
  {pages} {053613} (\bibinfo {year} {2008})}\BibitemShut {NoStop}%
\bibitem [{\citenamefont {Roy}\ and\ \citenamefont
  {Braunstein}(2008)}]{Roy2008}%
  \BibitemOpen
  \bibfield  {author} {\bibinfo {author} {\bibfnamefont {S.~M.}\ \bibnamefont
  {Roy}}\ and\ \bibinfo {author} {\bibfnamefont {Samuel~L.}\ \bibnamefont
  {Braunstein}},\ }\bibfield  {title} {\enquote {\bibinfo {title}
  {Exponentially enhanced quantum metrology},}\ }\href@noop {} {\bibfield
  {journal} {\bibinfo  {journal} {Phys. Rev. Lett.}\ }\textbf {\bibinfo
  {volume} {100}},\ \bibinfo {pages} {220501} (\bibinfo {year}
  {2008})}\BibitemShut {NoStop}%
\bibitem [{\citenamefont {Woolley}\ \emph {et~al.}(2008)\citenamefont
  {Woolley}, \citenamefont {Milburn},\ and\ \citenamefont
  {Caves}}]{Woolley2008}%
  \BibitemOpen
  \bibfield  {author} {\bibinfo {author} {\bibfnamefont {M.~J.}\ \bibnamefont
  {Woolley}}, \bibinfo {author} {\bibfnamefont {G.~J.}\ \bibnamefont
  {Milburn}}, \ and\ \bibinfo {author} {\bibfnamefont {C.~M}\ \bibnamefont
  {Caves}},\ }\bibfield  {title} {\enquote {\bibinfo {title} {Nonlinear quantum
  metrology using coupled nanomechanical resonators},}\ }\href@noop {}
  {\bibfield  {journal} {\bibinfo  {journal} {New J. Phys.}\ }\textbf {\bibinfo
  {volume} {10}},\ \bibinfo {pages} {125018} (\bibinfo {year}
  {2008})}\BibitemShut {NoStop}%
\bibitem [{\citenamefont {Napolitano}\ and\ \citenamefont
  {Mitchell}(2010)}]{Napolitano2010}%
  \BibitemOpen
  \bibfield  {author} {\bibinfo {author} {\bibfnamefont {M.}~\bibnamefont
  {Napolitano}}\ and\ \bibinfo {author} {\bibfnamefont {M.~W.}\ \bibnamefont
  {Mitchell}},\ }\bibfield  {title} {\enquote {\bibinfo {title} {Nonlinear
  metrology with a quantum interface},}\ }\href@noop {} {\bibfield  {journal}
  {\bibinfo  {journal} {New J. Phys.}\ }\textbf {\bibinfo {volume} {12}},\
  \bibinfo {pages} {093016} (\bibinfo {year} {2010})}\BibitemShut {NoStop}%
\bibitem [{\citenamefont {Rams}\ \emph {et~al.}(2018)\citenamefont {Rams},
  \citenamefont {Sierant}, \citenamefont {Dutta}, \citenamefont {Horodecki},\
  and\ \citenamefont {Zakrzewski}}]{Rams2018}%
  \BibitemOpen
  \bibfield  {author} {\bibinfo {author} {\bibfnamefont {M.~M.}\ \bibnamefont
  {Rams}}, \bibinfo {author} {\bibfnamefont {P.}~\bibnamefont {Sierant}},
  \bibinfo {author} {\bibfnamefont {O.}~\bibnamefont {Dutta}}, \bibinfo
  {author} {\bibfnamefont {P.}~\bibnamefont {Horodecki}}, \ and\ \bibinfo
  {author} {\bibfnamefont {J.}~\bibnamefont {Zakrzewski}},\ }\bibfield  {title}
  {\enquote {\bibinfo {title} {At the limits of criticality-based quantum
  metrology: Apparent super-heisenberg scaling revisited},}\ }\href@noop {}
  {\bibfield  {journal} {\bibinfo  {journal} {Phys. Rev. X}\ }\textbf {\bibinfo
  {volume} {8}},\ \bibinfo {pages} {021022} (\bibinfo {year}
  {2018})}\BibitemShut {NoStop}%
\bibitem [{\citenamefont {Napolitano}\ \emph {et~al.}(2011)\citenamefont
  {Napolitano}, \citenamefont {Koschorreck}, \citenamefont {Dubost},
  \citenamefont {Behbood}, \citenamefont {Sewell},\ and\ \citenamefont
  {Mitchell}}]{Napolitano2011}%
  \BibitemOpen
  \bibfield  {author} {\bibinfo {author} {\bibfnamefont {M.}~\bibnamefont
  {Napolitano}}, \bibinfo {author} {\bibfnamefont {M.}~\bibnamefont
  {Koschorreck}}, \bibinfo {author} {\bibfnamefont {B.}~\bibnamefont {Dubost}},
  \bibinfo {author} {\bibfnamefont {N.}~\bibnamefont {Behbood}}, \bibinfo
  {author} {\bibfnamefont {R.~J.}\ \bibnamefont {Sewell}}, \ and\ \bibinfo
  {author} {\bibfnamefont {M.~W.}\ \bibnamefont {Mitchell}},\ }\bibfield
  {title} {\enquote {\bibinfo {title} {Interaction-based quantum metrology
  showing scaling beyond the heisenberg limit},}\ }\href@noop {} {\bibfield
  {journal} {\bibinfo  {journal} {Nature}\ }\textbf {\bibinfo {volume} {471}},\
  \bibinfo {pages} {486} (\bibinfo {year} {2011})}\BibitemShut {NoStop}%
\bibitem [{\citenamefont {Borel}(1942)}]{Borel1942}%
  \BibitemOpen
  \bibfield  {author} {\bibinfo {author} {\bibfnamefont {E.}~\bibnamefont
  {Borel}},\ }\bibfield  {title} {\enquote {\bibinfo {title} {Sur l'emploi du
  th\'{e}or\`{e}me de bernoulli pour faciliter le calcul d'une infinit\'{e} de
  coefficients. application au probl\`{e}me de l'attente \`{a} un guichet},}\
  }\href@noop {} {\bibfield  {journal} {\bibinfo  {journal} {C. R. Acad. Sci.}\
  }\textbf {\bibinfo {volume} {214}},\ \bibinfo {pages} {452} (\bibinfo {year}
  {1942})}\BibitemShut {NoStop}%
\bibitem [{\citenamefont {Tanner}(1961)}]{Tanner1961}%
  \BibitemOpen
  \bibfield  {author} {\bibinfo {author} {\bibfnamefont {J.~C.}\ \bibnamefont
  {Tanner}},\ }\bibfield  {title} {\enquote {\bibinfo {title} {A derivation of
  the borel distribution},}\ }\href@noop {} {\bibfield  {journal} {\bibinfo
  {journal} {Biometrika}\ }\textbf {\bibinfo {volume} {48}},\ \bibinfo {pages}
  {222} (\bibinfo {year} {1961})}\BibitemShut {NoStop}%
\bibitem [{\citenamefont {Otter}(1949)}]{Otter1949}%
  \BibitemOpen
  \bibfield  {author} {\bibinfo {author} {\bibfnamefont {R.}~\bibnamefont
  {Otter}},\ }\bibfield  {title} {\enquote {\bibinfo {title} {The
  multiplicative process},}\ }\href@noop {} {\bibfield  {journal} {\bibinfo
  {journal} {Ann. Math. Stat.}\ }\textbf {\bibinfo {volume} {20}},\ \bibinfo
  {pages} {206} (\bibinfo {year} {1949})}\BibitemShut {NoStop}%
\bibitem [{\citenamefont {Haight}\ and\ \citenamefont
  {Breuer}(1960)}]{Haight1960}%
  \BibitemOpen
  \bibfield  {author} {\bibinfo {author} {\bibfnamefont {F.~A.}\ \bibnamefont
  {Haight}}\ and\ \bibinfo {author} {\bibfnamefont {M.~A.}\ \bibnamefont
  {Breuer}},\ }\bibfield  {title} {\enquote {\bibinfo {title} {The borel-tanner
  distribution},}\ }\href@noop {} {\bibfield  {journal} {\bibinfo  {journal}
  {Biometrika}\ }\textbf {\bibinfo {volume} {47}},\ \bibinfo {pages} {143}
  (\bibinfo {year} {1960})}\BibitemShut {NoStop}%
\bibitem [{\citenamefont {Foss}\ \emph {et~al.}(2013)\citenamefont {Foss},
  \citenamefont {Korshunov},\ and\ \citenamefont {Zachary}}]{Foss2013}%
  \BibitemOpen
  \bibfield  {author} {\bibinfo {author} {\bibfnamefont {S.}~\bibnamefont
  {Foss}}, \bibinfo {author} {\bibfnamefont {D.}~\bibnamefont {Korshunov}}, \
  and\ \bibinfo {author} {\bibfnamefont {S.}~\bibnamefont {Zachary}},\
  }\href@noop {} {\emph {\bibinfo {title} {An Introduction to Heavy-Tailed and
  Subexponential Distributions}}}\ (\bibinfo  {publisher} {Springer-Verlag New
  York},\ \bibinfo {year} {2013})\BibitemShut {NoStop}%
\bibitem [{\citenamefont {Zhang}\ \emph {et~al.}(2013)\citenamefont {Zhang},
  \citenamefont {Jin}, \citenamefont {Cao}, \citenamefont {Liu},\ and\
  \citenamefont {Fan}}]{Zhang2013}%
  \BibitemOpen
  \bibfield  {author} {\bibinfo {author} {\bibfnamefont {Y.~R.}\ \bibnamefont
  {Zhang}}, \bibinfo {author} {\bibfnamefont {G.~R.}\ \bibnamefont {Jin}},
  \bibinfo {author} {\bibfnamefont {J.~P.}\ \bibnamefont {Cao}}, \bibinfo
  {author} {\bibfnamefont {W.~M.}\ \bibnamefont {Liu}}, \ and\ \bibinfo
  {author} {\bibfnamefont {H.}~\bibnamefont {Fan}},\ }\bibfield  {title}
  {\enquote {\bibinfo {title} {Unbounded quantum \text{Fisher} information in
  two-path interferometry with finite photon number},}\ }\href@noop {}
  {\bibfield  {journal} {\bibinfo  {journal} {J. Phys. A: Math. Theor.}\
  }\textbf {\bibinfo {volume} {46}},\ \bibinfo {pages} {035302} (\bibinfo
  {year} {2013})}\BibitemShut {NoStop}%
\bibitem [{\citenamefont {Pezz{\`e}}\ and\ \citenamefont
  {Smerzi}(2008)}]{Pezze2008}%
  \BibitemOpen
  \bibfield  {author} {\bibinfo {author} {\bibfnamefont {L.}~\bibnamefont
  {Pezz{\`e}}}\ and\ \bibinfo {author} {\bibfnamefont {A.}~\bibnamefont
  {Smerzi}},\ }\bibfield  {title} {\enquote {\bibinfo {title}
  {Mach-\text{Zehnder} interferometry at the heisenberg limit with coherent and
  squeezed-vacuum light},}\ }\href@noop {} {\bibfield  {journal} {\bibinfo
  {journal} {Phys. Rev. Lett.}\ }\textbf {\bibinfo {volume} {100}},\ \bibinfo
  {pages} {073601} (\bibinfo {year} {2008})}\BibitemShut {NoStop}%
\bibitem [{\citenamefont {Giovannetti}\ \emph {et~al.}(2004)\citenamefont
  {Giovannetti}, \citenamefont {Lloyd},\ and\ \citenamefont
  {Maccone}}]{Giovannetti2004}%
  \BibitemOpen
  \bibfield  {author} {\bibinfo {author} {\bibfnamefont {V.}~\bibnamefont
  {Giovannetti}}, \bibinfo {author} {\bibfnamefont {S.}~\bibnamefont {Lloyd}},
  \ and\ \bibinfo {author} {\bibfnamefont {L.}~\bibnamefont {Maccone}},\
  }\bibfield  {title} {\enquote {\bibinfo {title} {Quantum-enhanced
  measurements: beating the standard quantum limit},}\ }\href@noop {}
  {\bibfield  {journal} {\bibinfo  {journal} {Science}\ }\textbf {\bibinfo
  {volume} {306}},\ \bibinfo {pages} {1330} (\bibinfo {year}
  {2004})}\BibitemShut {NoStop}%
\bibitem [{\citenamefont {Hradil}\ and\ \citenamefont
  {Shapiro}(1992)}]{Hradil1992a}%
  \BibitemOpen
  \bibfield  {author} {\bibinfo {author} {\bibfnamefont {Z.}~\bibnamefont
  {Hradil}}\ and\ \bibinfo {author} {\bibfnamefont {J.~H.}\ \bibnamefont
  {Shapiro}},\ }\bibfield  {title} {\enquote {\bibinfo {title} {Quantum phase
  measurements with infinite peak-likelihood and zero phase information},}\
  }\href@noop {} {\bibfield  {journal} {\bibinfo  {journal} {Quantum Opt.}\
  }\textbf {\bibinfo {volume} {4}},\ \bibinfo {pages} {31} (\bibinfo {year}
  {1992})}\BibitemShut {NoStop}%
\bibitem [{\citenamefont {Hradil}(1992)}]{Hradil1992b}%
  \BibitemOpen
  \bibfield  {author} {\bibinfo {author} {\bibfnamefont {Z.}~\bibnamefont
  {Hradil}},\ }\bibfield  {title} {\enquote {\bibinfo {title} {Performance
  measures of quantum-phase measurement},}\ }\href@noop {} {\bibfield
  {journal} {\bibinfo  {journal} {Phys. Rev. A}\ }\textbf {\bibinfo {volume}
  {46}},\ \bibinfo {pages} {R2217} (\bibinfo {year} {1992})}\BibitemShut
  {NoStop}%
\bibitem [{\citenamefont {Braunstein}\ \emph {et~al.}(1992)\citenamefont
  {Braunstein}, \citenamefont {Lane},\ and\ \citenamefont
  {Caves}}]{Braunstein1992a}%
  \BibitemOpen
  \bibfield  {author} {\bibinfo {author} {\bibfnamefont {S.~L.}\ \bibnamefont
  {Braunstein}}, \bibinfo {author} {\bibfnamefont {A.~S.}\ \bibnamefont
  {Lane}}, \ and\ \bibinfo {author} {\bibfnamefont {C.~M.}\ \bibnamefont
  {Caves}},\ }\bibfield  {title} {\enquote {\bibinfo {title}
  {Maximum-likelihood analysis of multiple quantum phase measurements},}\
  }\href@noop {} {\bibfield  {journal} {\bibinfo  {journal} {Phys. Rev. Lett.}\
  }\textbf {\bibinfo {volume} {69}},\ \bibinfo {pages} {2153} (\bibinfo {year}
  {1992})}\BibitemShut {NoStop}%
\bibitem [{\citenamefont {Braunstein}(1992)}]{Braunstein1992b}%
  \BibitemOpen
  \bibfield  {author} {\bibinfo {author} {\bibfnamefont {S.~L.}\ \bibnamefont
  {Braunstein}},\ }\bibfield  {title} {\enquote {\bibinfo {title} {Quantum
  limits on precision measurements of phase},}\ }\href@noop {} {\bibfield
  {journal} {\bibinfo  {journal} {Phys. Rev. Lett.}\ }\textbf {\bibinfo
  {volume} {69}},\ \bibinfo {pages} {3598} (\bibinfo {year}
  {1992})}\BibitemShut {NoStop}%
\bibitem [{\citenamefont {Lane}\ \emph {et~al.}(1993)\citenamefont {Lane},
  \citenamefont {Braunstein},\ and\ \citenamefont {Caves}}]{Lane1993}%
  \BibitemOpen
  \bibfield  {author} {\bibinfo {author} {\bibfnamefont {A.~S.}\ \bibnamefont
  {Lane}}, \bibinfo {author} {\bibfnamefont {S.~L.}\ \bibnamefont
  {Braunstein}}, \ and\ \bibinfo {author} {\bibfnamefont {C.~M.}\ \bibnamefont
  {Caves}},\ }\bibfield  {title} {\enquote {\bibinfo {title} {Maximum-
  likelihood statistics of multiple quantum phase measurements},}\ }\href@noop
  {} {\bibfield  {journal} {\bibinfo  {journal} {Phys. Rev. A}\ }\textbf
  {\bibinfo {volume} {47}},\ \bibinfo {pages} {1667} (\bibinfo {year}
  {1993})}\BibitemShut {NoStop}%
\bibitem [{\citenamefont {Demkowicz-Dobrza{\' n}ski}\ \emph
  {et~al.}(2012)\citenamefont {Demkowicz-Dobrza{\' n}ski}, \citenamefont
  {Ko{\l}ody{\' n}ski},\ and\ \citenamefont
  {M.G\c{t}\u{a}}}]{Demkowicz-Dobrzanski2012}%
  \BibitemOpen
  \bibfield  {author} {\bibinfo {author} {\bibfnamefont {R.}~\bibnamefont
  {Demkowicz-Dobrza{\' n}ski}}, \bibinfo {author} {\bibfnamefont
  {J.}~\bibnamefont {Ko{\l}ody{\' n}ski}}, \ and\ \bibinfo {author}
  {\bibnamefont {M.G\c{t}\u{a}}},\ }\bibfield  {title} {\enquote {\bibinfo
  {title} {The elusive heisenberg limit in quantum-enhanced metrology},}\
  }\href@noop {} {\bibfield  {journal} {\bibinfo  {journal} {Nat. Commun.}\
  }\textbf {\bibinfo {volume} {3}},\ \bibinfo {pages} {1063} (\bibinfo {year}
  {2012})}\BibitemShut {NoStop}%
\bibitem [{\citenamefont {Pezz{\' e}}(2013)}]{Pezze2013}%
  \BibitemOpen
  \bibfield  {author} {\bibinfo {author} {\bibfnamefont {L.}~\bibnamefont
  {Pezz{\' e}}},\ }\bibfield  {title} {\enquote {\bibinfo {title}
  {Sub-heisenberg phase uncertainties},}\ }\href@noop {} {\bibfield  {journal}
  {\bibinfo  {journal} {Phys. Rev. A.}\ }\textbf {\bibinfo {volume} {88}},\
  \bibinfo {pages} {060101(R)} (\bibinfo {year} {2013})}\BibitemShut {NoStop}%
\bibitem [{\citenamefont {Pezz{\' e}}\ \emph {et~al.}(2015)\citenamefont
  {Pezz{\' e}}, \citenamefont {Hyllus},\ and\ \citenamefont
  {Smerzi}}]{Pezze2015}%
  \BibitemOpen
  \bibfield  {author} {\bibinfo {author} {\bibfnamefont {L.}~\bibnamefont
  {Pezz{\' e}}}, \bibinfo {author} {\bibfnamefont {P.}~\bibnamefont {Hyllus}},
  \ and\ \bibinfo {author} {\bibfnamefont {A.}~\bibnamefont {Smerzi}},\
  }\bibfield  {title} {\enquote {\bibinfo {title} {Phase-sensitivity bounds for
  two-mode interferometers},}\ }\href@noop {} {\bibfield  {journal} {\bibinfo
  {journal} {Phys. Rev. A}\ }\textbf {\bibinfo {volume} {91}},\ \bibinfo
  {pages} {032103} (\bibinfo {year} {2015})}\BibitemShut {NoStop}%
\bibitem [{\citenamefont {Streltsov}\ \emph {et~al.}(2017)\citenamefont
  {Streltsov}, \citenamefont {Adesso},\ and\ \citenamefont
  {Plenio}}]{Streltsov2017}%
  \BibitemOpen
  \bibfield  {author} {\bibinfo {author} {\bibfnamefont {A.}~\bibnamefont
  {Streltsov}}, \bibinfo {author} {\bibfnamefont {G.}~\bibnamefont {Adesso}}, \
  and\ \bibinfo {author} {\bibfnamefont {M.~B.}\ \bibnamefont {Plenio}},\
  }\bibfield  {title} {\enquote {\bibinfo {title} {Quantum coherence as a
  resource},}\ }\href@noop {} {\bibfield  {journal} {\bibinfo  {journal} {Rev.
  Mod. Phys.}\ }\textbf {\bibinfo {volume} {89}},\ \bibinfo {pages} {041003}
  (\bibinfo {year} {2017})}\BibitemShut {NoStop}%
\bibitem [{\citenamefont {Giorda}\ and\ \citenamefont
  {Allegra}(2018)}]{Giorda2018}%
  \BibitemOpen
  \bibfield  {author} {\bibinfo {author} {\bibfnamefont {P.}~\bibnamefont
  {Giorda}}\ and\ \bibinfo {author} {\bibfnamefont {M.}~\bibnamefont
  {Allegra}},\ }\bibfield  {title} {\enquote {\bibinfo {title} {Coherence in
  quantum estimation},}\ }\href@noop {} {\bibfield  {journal} {\bibinfo
  {journal} {J. Phys. A: Math. Theor.}\ }\textbf {\bibinfo {volume} {51}},\
  \bibinfo {pages} {025302} (\bibinfo {year} {2018})}\BibitemShut {NoStop}%
\bibitem [{\citenamefont {Tan}\ \emph {et~al.}(2018)\citenamefont {Tan},
  \citenamefont {Choi}, \citenamefont {Kwon},\ and\ \citenamefont
  {Jeong}}]{Tan2018}%
  \BibitemOpen
  \bibfield  {author} {\bibinfo {author} {\bibfnamefont {K.~C.}\ \bibnamefont
  {Tan}}, \bibinfo {author} {\bibfnamefont {S.}~\bibnamefont {Choi}}, \bibinfo
  {author} {\bibfnamefont {H.}~\bibnamefont {Kwon}}, \ and\ \bibinfo {author}
  {\bibfnamefont {H.}~\bibnamefont {Jeong}},\ }\bibfield  {title} {\enquote
  {\bibinfo {title} {Coherence, quantum fisher information, superradiance, and
  entanglement as interconvertible resources},}\ }\href@noop {} {\bibfield
  {journal} {\bibinfo  {journal} {Phys. Rev. A}\ }\textbf {\bibinfo {volume}
  {97}},\ \bibinfo {pages} {052304} (\bibinfo {year} {2018})}\BibitemShut
  {NoStop}%
\bibitem [{\citenamefont {Kwon}\ \emph {et~al.}(2018)\citenamefont {Kwon},
  \citenamefont {Tan}, \citenamefont {Choi},\ and\ \citenamefont
  {Jeong}}]{Kwon2018}%
  \BibitemOpen
  \bibfield  {author} {\bibinfo {author} {\bibfnamefont {H.}~\bibnamefont
  {Kwon}}, \bibinfo {author} {\bibfnamefont {K.~C.}\ \bibnamefont {Tan}},
  \bibinfo {author} {\bibfnamefont {S.}~\bibnamefont {Choi}}, \ and\ \bibinfo
  {author} {\bibfnamefont {H.}~\bibnamefont {Jeong}},\ }\bibfield  {title}
  {\enquote {\bibinfo {title} {Quantum fisher information on its own is not a
  valid measure of the coherence},}\ }\href@noop {} {\bibfield  {journal}
  {\bibinfo  {journal} {Results Phys.}\ }\textbf {\bibinfo {volume} {9}},\
  \bibinfo {pages} {1594} (\bibinfo {year} {2018})}\BibitemShut {NoStop}%
\bibitem [{\citenamefont {Lee}\ \emph {et~al.}(2016)\citenamefont {Lee},
  \citenamefont {Lee}, \citenamefont {Lee},\ and\ \citenamefont
  {Nha}}]{Lee2016}%
  \BibitemOpen
  \bibfield  {author} {\bibinfo {author} {\bibfnamefont {S.-Y.}\ \bibnamefont
  {Lee}}, \bibinfo {author} {\bibfnamefont {C.-W.}\ \bibnamefont {Lee}},
  \bibinfo {author} {\bibfnamefont {J.}~\bibnamefont {Lee}}, \ and\ \bibinfo
  {author} {\bibfnamefont {H.}~\bibnamefont {Nha}},\ }\bibfield  {title}
  {\enquote {\bibinfo {title} {Quantum phase estimation using path-symmetric
  entangled states},}\ }\href@noop {} {\bibfield  {journal} {\bibinfo
  {journal} {Sci. Rep.}\ }\textbf {\bibinfo {volume} {6}},\ \bibinfo {pages}
  {30306} (\bibinfo {year} {2016})}\BibitemShut {NoStop}%
\bibitem [{\citenamefont {Roy}(2019)}]{Roy2019}%
  \BibitemOpen
  \bibfield  {author} {\bibinfo {author} {\bibfnamefont {S.}~\bibnamefont
  {Roy}},\ }\bibfield  {title} {\enquote {\bibinfo {title} {Fundamental noisy
  multiparameter quantum bounds},}\ }\href@noop {} {\bibfield  {journal}
  {\bibinfo  {journal} {Sci. Rep.}\ }\textbf {\bibinfo {volume} {9}},\ \bibinfo
  {pages} {1038} (\bibinfo {year} {2019})}\BibitemShut {NoStop}%
\end{thebibliography}%

\end{document}